\documentclass[11pt, a4paper]{scrartcl}

\usepackage{latexsym}
\usepackage{amssymb}
\usepackage[nonamelimits]{amsmath}
\usepackage[authoryear, round]{natbib}

\usepackage{color}
\definecolor{myred}{rgb}{0.5,0,0}
\definecolor{myblue}{rgb}{0,0,0.75}
\definecolor{mygreen}{rgb}{0,0.5,0}

\usepackage{ifpdf}

\ifpdf
   \usepackage[pdftex]{graphicx}
   \usepackage[pdftex,
               pdftitle={The art of probability-of-default curve calibration},
               pdfsubject={},
               pdfkeywords={Probability of default, calibration, likelihood ratio, Bayes' formula, rating profile, binary classification},
               pdfauthor={Dirk Tasche},
               pdfstartview=FitR,
               breaklinks=true,
               colorlinks=true,
               citecolor=myred,
               linkcolor=myblue,
               urlcolor=mygreen]{hyperref}
\else
   \usepackage[dvips]{graphicx}
   \usepackage{hyperref}
   \usepackage{rotating}
\fi

\newtheorem{theorem}{Theorem}[section]

\newtheorem{proposition}[theorem]{Proposition}

\newtheorem{assumption}[theorem]{Assumption}

\numberwithin{equation}{section}

\title{The art of probability-of-default curve calibration}

\author{%
Dirk Tasche\thanks{E-mail: dirk.tasche@gmx.net\newline
The author currently works at the Prudential Regulation Authority (a
division of the Bank of England). 
The opinions expressed in this paper are those of the author 
and do not necessarily reflect views of the Bank of England.\newline
The author is grateful to two anonymous referees whose suggestions significantly
improved the paper.}}

\date{First version: December 15, 2012\\
This version: November 26, 2013}

\begin{document}
\maketitle

\begin{abstract}
PD curve calibration refers to the transformation of a set of rating grade level 
probabilities of default (PDs) to another average PD level that is 
determined by a change of the underlying portfolio-wide PD. This paper 
presents a framework that allows to explore a variety of calibration
approaches and the conditions under which they are fit for purpose. 
We test the approaches discussed by applying them to publicly available 
datasets of agency rating and default statistics
that can be considered typical for the scope of application of the approaches.
We show that the popular `scaled PDs' approach  is 
theoretically questionable and
identify an alternative calibration approach (`scaled likelihood ratio') that is both theoretically sound
and performs better on the test datasets.
\end{abstract}


\section{Introduction}
\label{se:intro}

The best way to understand the subject of this paper is to have a glance at table~\ref{tab:problem}
on page~\pageref{tab:problem} that illustrates the problem studied. Table~\ref{tab:problem} shows 
the grade-level and portfolio-wide default rates (third column) that were observed in 2009 for
S\&P-rated corporate entities together with the rating frequencies that were observed at 
the beginning of 2009 (second column) and at the beginning of 2010 (fourth column). The question
marks in the fifth column indicate the question this paper is intended to answer: How can
grade-level default rates for a future time period be forecast on the basis of observations
from an earlier period and the known rating profile at the beginning of the future period?

The question mark in the lower right corner of table~\ref{tab:problem} indicates that
we investigate this question both under the assumption that an independent forecast of the future
portfolio-wide default rate is known and under the assumption that also the future
portfolio-wide default rate has to be forecast. 
  
\begin{table}[t!p]
\caption{S\&P rating frequencies (\%) and default rates (\%) in 2009 and rating frequencies in 2010.
Sources: \citet{S&P2010}, tables 51 to 53, and \citet{S&P2011}, tables 50 to 52.}
\label{tab:problem}
\begin{center}
\begin{tabular}{|l||r|r||r|c|}
\hline
& \multicolumn{2}{|c||}{2009} & \multicolumn{2}{|c|}{2010} \\ \hline
Rating & Frequency & Default rate & Frequency & Default rate \\ \hline\hline
AAA  & 1.38 & 0 & 1.3 & ?  \\ \hline
AA+  & 0.63 & 0 & 0.45 & ?  \\ \hline
AA  & 3.21 & 0 & 2.59 & ?  \\ \hline
AA-  & 4.18 & 0 & 3.78 & ?   \\ \hline
A+  & 5.8 & 0.29 & 6.39 & ?   \\ \hline
A  & 8.7 & 0.39 & 8.58 & ?  \\ \hline
A-  & 9.32 & 0 & 9.56 & ?  \\ \hline
BBB+  & 8.5 & 0.4 & 8.28 & ?  \\ \hline
BBB  & 9.23 & 0.18 & 10.56 & ?  \\ \hline
BBB-  & 7.83 & 1.09 & 7.79 & ? \\ \hline
BB+  & 4.54 & 0 & 4.6 & ? \\ \hline
BB  & 5.03 & 1.02 & 5 & ? \\ \hline
BB-  & 7.53 & 0.91 & 6.86 & ? \\ \hline
B+  & 7.47 & 5.48 & 7.12 & ? \\ \hline
B  & 8.23 & 9.96 & 7.9 & ? \\ \hline
B-  & 5.17 & 17.16 & 5.25 & ? \\ \hline
CCC-C  & 3.24 & 48.42 & 3.98 & ? \\ \hline\hline
All  & 100 & 3.99 & 100 & 1.14? \\ \hline
\end{tabular}
\end{center}
\end{table}
We call a forecast of grade-level default rates a \emph{PD curve.}
The problem we study in this paper is made more complicated by the fact that for economic reasons
PD curves are subject to the constraints that they need to be monotonic and positive -- although
table~\ref{tab:problem} shows that this is not necessarily true for empirically observed default
rates. 

The scope of the concepts and approaches described in this paper is not limited to
data from rating agencies but covers any rating system for which data from an 
\emph{estimation period} is available. It should, however, be noted that
the focus in this paper is on grade-level default rate forecasts while
the problem of forecasting the unconditional (or portfolio-wide) default rate
is not considered. Forecasting the unconditional default rate is
an econometric problem that is beyond the scope of this paper 
\citep[see][for an example of how to approach this problem]{Engelmann&Porath2012}.

This paper appears to be almost unique in that it solely deals with the calibration or
recalibration of PD curves. Calibration of PD curves is a topic that is often mentioned
in the literature but mostly only as one aspect of the more general subject of
rating model development. For instance, \citet{Falkenstein&Boral&Carty} 
deployed the approach that is called `scaled PDs' in this paper
without any comment about why they considered it appropriate. There are, however,
some authors who devoted complete articles or book sections to PD curve calibration.
\Citet{VanDerBurgt} suggested a predecessor of the technique that is called quasi moment matching
(QMM) in this paper. \citet[][chapter~4]{Bohn&Stein} discussed the 
conceptual and practical differences between the `scaled PDs' and `invariant 
likelihood ratio' approaches. More recently, \citet{Konrad.Diss} investigated in some detail
the interplay between the calibration and the discriminatory power of rating models.

In this paper, we revisit the concept of two calibration steps as used by \citet{Bohn&Stein}.
According to \citet{Bohn&Stein} the two steps are a consequence of the fact that usually the first calibration of
a rating model is conducted on a training sample in which the proportion of good and 
bad might not be representative of the live portfolio. The second calibration step, therefore,
is needed to adjust the calibration to the right proportion of good and bad.

We argue more generally that the two steps actually relate to different time periods (the estimation and
the forecast periods) which both can be described by the same type of model. This
view encompasses both the situation where a new rating model is calibrated and the situation where
an existing rating model undergoes a -- possibly periodic -- recalibration. 
The estimation period is used to estimate the model components that are assumed to be invariant 
(i.e.\ unchanged) or in a specific way transformed between the estimation and the forecast periods.
Calibration approaches for the forecast period are essentially determined by 
the assumptions of invariance between the periods.

Specifically, the model estimation in the estimation period involves smoothing of the 
observed default rates in order to create a positive and monotonic PD curve. For this purpose
we apply quasi moment matching (QMM) the details of which are described in appendix~\ref{se:app}.

When in the following we investigate different invariance assumptions that can be made for the forecast period the
basic idea is always that the rating system's discriminatory power is the same or nearly the
same both
in the estimation and forecast periods. However, discriminatory power can technically be 
expressed in a number of different ways that correspond to invariance assumptions with
a range of different implications. This is why we first study in section~\ref{se:mecha} in some detail the model
components that are related to invariance assumptions:
\begin{itemize}
	\item Unconditional rating distribution (profile).
	\item Conditional (on default and survival) rating distributions (profiles).
	\item Unconditional PD.
	\item PD curve (grade-level PDs, i.e.~PDs conditional on rating grades).
	\item Accuracy ratio (as a measure of discriminatory power).
	\item Likelihood ratio.
\end{itemize}
In particular, we derive a new result (theorem~\ref{th:unique}) on
	the characterisation of the joint distribution of a borrower's rating
	at the beginning of the observation period and his solvency state at the end
	of the period by unconditional rating profile and likelihood ratio.
	
Then, in section~\ref{se:calibration}, we 
look at different calibration approaches (which
may be characterised by invariance assumptions).
The suitability of the approaches described depends strongly upon
	what data (e.g.\ the unconditional rating profile) can be observed
	at the time when the forecast exercise takes place. We therefore discuss
	the different possibilities and assumptions in some detail and examine the performance of the approaches
	with a real data example. The example is based on the S\&P data from table~\ref{tab:problem} 
	which is presented in more detail in section~\ref{se:data}. 
	
	In particular, the example in section~\ref{se:calibration} suggests that the popular 
	'scaled PDs' approach (corresponding to the assumption
	of an invariant shape of the PD curve)  is both theoretically
	questionable and not very well performing on the example dataset.
	Two other approaches (`invariant AR' and `scaled likelihood ratio')
	appear to be theoretically sound and better performing when deployed for
	the numerical example.
	
However, as the S\&P dataset is small the example provides anecdotal evidence only. Its suggestions are therefore
backtested and qualified in section~\ref{se:backtest}. The -- still rather limited -- backtest confirms that
the `scaled likelihood ratio' approach performs better than the 'scaled PDs' approach.
In contrast, the `invariant AR' approach is found to be underperforming in the backtest.


\section{Data and context}
\label{se:data}

The numerical examples in section~\ref{se:calibration} in this paper are based on the S\&P rating  and default
statistics for all corporates as
presented in table \ref{tab:all} on page~\pageref{tab:all}. Only with 
their 2009 default data report \citet{S&P2010} began to
make information on modified-grade level issuer numbers readily available.
Without issuer numbers, however, there is not sufficient information
to calculate rating profiles and conduct goodness-of-fit tests for
rating profiles because such tests typically require the
occupation frequencies (i.e.\ the numbers of issuers in each of
the rating grades) as input.
This explains why we only look at default statistics from 2009 onwards. 

For the purposes of this paper, data from 
Moody's is less suitable because Moody's do not provide issuer
numbers at alphanumeric grade level and estimate default rates 
in a way that makes it impossible to infer exact grade-level numbers
of defaults \citep[see][for details of the estimation approach]{Hamilton&Cantor}.
Therefore, in order to work with the publicly available \cite{Moodys2013} data, one
has to make assumptions that are likely to make the results less reliable. 
That is why, in section~\ref{se:backtest}, we use Moody's default and rating data
only for backtesting and qualifying the observations from section~\ref{se:calibration}. 

\begin{table}[t!p]
\caption{S\&P's corporate 
ratings, defaults and default rates (DR, \%) in 2009, 2010 and 2011. Sources:
\citet[][tables 51 to 53]{S&P2010}, \citet[][tables 50 to 52]{S&P2011}, 
\citet[][tables 50 to 52]{S&P2012}.}
\label{tab:all}
\begin{center}
\begin{tabular}{|l||c|c|c||c|c|c||c|c|c|}
\hline
 & \multicolumn{3}{c||}{2009} & \multicolumn{3}{c||}{2010} & \multicolumn{3}{c|}{2011}\\ \hline
Rating grade & rated & defaults & DR & rated & defaults & DR & rated & defaults & DR \\ \hline \hline
AAA & 81 & 0 & 0.00 & 72 & 0 & 0.00 & 51 & 0 & 0.00\\ \hline
AA+ & 37 & 0 & 0.00 & 25 & 0 & 0.00 & 36 & 0 & 0.00\\ \hline
AA & 188 & 0 & 0.00 & 143 & 0 & 0.00 & 120 & 0 & 0.00\\ \hline
AA- & 245 & 0  & 0.00 & 209 & 0 & 0.00 & 207 & 0 & 0.00 \\ \hline
A+ & 340 & 1 & 0.29 & 353 & 0 & 0.00 & 357 & 0 & 0.00\\ \hline
A & 510 & 2 & 0.39 & 474 & 0 & 0.00 & 470 & 0 & 0.00\\ \hline
A- & 546 & 0 & 0.00 & 528 & 0 & 0.00 & 560 & 0 & 0.00\\ \hline
BBB+ & 498 & 2 & 0.40 & 457 & 0 & 0.00 & 473 & 0 & 0.00\\ \hline
BBB & 541 & 1 & 0.18 & 583 & 0 & 0.00 & 549 & 0 & 0.00\\ \hline
BBB- & 459 & 5 & 1.09 & 430 & 0 & 0.00 & 508 & 1 & 0.20\\ \hline
BB+ & 266 & 0 & 0.00 & 254 & 2 & 0.79 & 260 & 0 & 0.00\\ \hline
BB & 295 & 3 & 1.02 & 276 & 1 & 0.36 & 319 & 0 & 0.00\\ \hline
BB- & 441 & 4 & 0.91 & 379 & 2 & 0.53 & 403 & 0 & 0.00\\ \hline
B+ & 438 & 24 & 5.48 & 393 & 0 & 0.00 & 509 & 2 & 0.39\\ \hline
B & 482 & 48 & 9.96 & 436 & 3 & 0.69  & 586 & 7 & 1.19\\ \hline
B- & 303 & 52 & 17.16 & 290 & 6 & 2.07 & 301 & 12 & 3.99\\ \hline
CCC-C & 190 & 92 & 48.42 & 220 & 49 & 22.27 & 138 & 22 & 15.94\\ \hline\hline
All & 5860 & 234 & 3.99 & 5522 & 63 & 1.14 & 5847 & 44 & 0.75\\ \hline
\end{tabular}
\end{center}
\end{table}


\subsection{Observations on the data}
\label{se:observations} 

S\&P's all corporates default statistics (table~\ref{tab:all})
represent an example of an interesting, somewhat problematic dataset because
it includes some instances of \emph{inversions} of observed default rates. 
`Inversion of default rates' means that the default rate observed for a better
rating grade is higher than the default rate of the adjacent worse rating grade.

The default rate columns of table~\ref{tab:all} show that
the corporate grade-level default rates recorded by S\&P
for 2009, 2010 and 2011 
increase in general with deteriorating credit quality as one would expect. However,
there are a number of `inversions' in all the default rate columns of the table,
i.e.\ there are some counter-intuitive examples of adjacent rating grades where the less risky
grade has a higher default rate than the adjacent riskier grade. Notable for this phenomenon
is, in particular, the pair of BBB- and BB+ in 2009 with 1.09\% defaults in BBB- and
0\% defaults in BB+.

Should we conclude from the existence of such inversions that there is
a problem with the rank-ordering capacity of the rating methodology? The long-run
average grade-level default rates reported by \citet[][table~23]{S&P2013}
suggest that the observation of default rate inversions as in table~\ref{tab:all}
might be an exception. By Fisher's exact test \citep[][Example 8.3.30]{FisherExact, Casella&Berger}
this explanation can be verified. 

\begin{table}[t!p]
\caption{S\&P rating profiles for
corporates at the beginning of 2009, 2010 and 2011. Sources:
\citet[][tables 51 to 53]{S&P2010}, \citet[][tables 50 to 52]{S&P2011}, 
\citet[][tables 50 to 52]{S&P2012} and own calculations. All
values in \%.}
\label{tab:profiles}
\begin{center}
\begin{tabular}{|l||c|c|c|}
\hline
Rating grade & 2009 & 2010 & 2011 \\ \hline \hline
AAA & 1.38 & 1.30 & 0.87  \\ \hline
AA+ & 0.63 & 0.45 & 0.62  \\ \hline
AA & 3.21 & 2.59 & 2.05  \\ \hline
AA- & 4.18 & 3.78 & 3.54  \\ \hline
A+ & 5.80 & 6.39 & 6.11  \\ \hline
A & 8.70 & 8.58 & 8.04  \\ \hline
A- & 9.32 & 9.56 & 9.58  \\ \hline
BBB+ & 8.50 & 8.28 & 8.09 \\ \hline
BBB & 9.23 & 10.56 & 9.39  \\ \hline
BBB- & 7.83 & 7.79 & 8.69 \\ \hline
BB+ & 4.54 & 4.60 & 4.45 \\ \hline
BB & 5.03 & 5.00 & 5.46 \\ \hline
BB- & 7.53 & 6.86 & 6.89 \\ \hline
B+ & 7.47 & 7.12 & 8.71 \\ \hline
B & 8.23 & 7.90 & 10.02 \\ \hline
B- & 5.17 & 5.25 & 5.15 \\ \hline
CCC-C & 3.24 & 3.98 & 2.36  \\ \hline\hline
All & 100.00 & 100.00 & 100.00  \\ \hline
\end{tabular}
\end{center}
\end{table}

A question of similar importance for the estimation of PD curves
is the question of whether or not the unconditional (or all-portfolio) rating profile
(i.e.\ the distribution of the rating grades)
of a portfolio can be assumed to be unchanged over time.
Table \ref{tab:profiles} on page~\pageref{tab:profiles} shows the unconditional rating profiles
of the corporate entities 
for the three years of S\&P data used in this paper. It appears 
from the percentages that the profiles vary significantly during
the three years  even if random differences are
ignored.  Pearson's $\chi^2$ test for count data \citep{Pearson1900, VanDerVaart} can be used
to test these observations and also to assess the
accuracy of the forecast approaches discussed in the remainder of the paper.


\subsection{Consequences for the calibration of PD curves}
\label{se:consequences}

From the observations in section \ref{se:observations}  we can draw two conclusions:
\begin{itemize}
	\item Forcing monotonicity of estimated PD curves can make sense 
if it is justified by statistical tests or long-run average evidence. 
	\item In general, we cannot assume that the rating profile of a portfolio
does not change over time, even if random fluctuation is ignored. 
However, this assumption can be verified or proven wrong with
statistical tests. Depending
on the outcome of the tests there are different options for the
estimation of PD curves. This will be discussed in detail in section~\ref{se:calibration}.
\end{itemize}
Although never a default of an AAA-rated
corporate was observed within one year after having been
rated AAA \citep{Moodys2013, S&P2013} 
we nonetheless try and infer a positive one-year
PD for AAA. 
This is why in the following we restrict ourselves to only deploy
PD curve estimation approaches that guarantee to deliver positive PDs for
all rating grades.  

On the basis of the data presented in this section, it is also worthwhile to 
clarify precisely the concept of a two-step (or two-period) approach to the calibration of 
a rating model as mentioned by \citet{Bohn&Stein}: The first period is the \emph{estimation} period, 
the second period is the \emph{calibration and forecast} period. 
The two periods are determined by their start and end dates and the observation and estimation horizon:
\begin{itemize}
	\item $h$ is the horizon for the PD estimation, i.e.\ a borrower's PD at date $T$ gives the probability
	that the borrower will default between $T$ and $T+h$. 
	\item The start date $T_0$ of the estimation period is a date in the past. 
	\item $T_1 \ge T_0+h$ is the date when the calibration or recalibration of the rating model takes place. The 
	calibration is for the current portfolio of borrowers whose ratings at $T_1$ should be known but whose
	future default status at $T_1+h$ is still unknown.
	\item The end date $T_2 = T_1+h$ of the forecast period is in the future. 
	Then the default status of the borrowers in the current portfolio will be known. 
\end{itemize}
With regard to the two-period concept for calibration, for the remainder of the paper we make the
following crucial assumptions: 
\begin{itemize}
	\item For the sample as of date $T_0$ everything is known:
\begin{itemize}
	\item The unconditional rating profile at $T_0$, 
	\item the conditional rating profiles (i.e.\ conditional on
	default and conditional on survival respectively) at $T_0$, 
	\item the unconditional (base) PD (estimated by the observed unconditional default rate) for the time interval between $T_0$ and $T_0 + h$, 
	\item	the conditional PDs (i.e.\ conditional on the rating grades, estimated by smoothing the observed grade-level default rates) at $T_0$ 
	for the time interval between $T_0$ and $T_0 + h$.
\end{itemize}
	\item At date $T_1$ could be known:
\begin{itemize}
	\item The unconditional rating profile.
	\item A forecast of the unconditional (base) default rate. 
	In general this will be different from the unconditional PD for the estimation period
	between $T_0$ and $T_0+h$.
\end{itemize}
\end{itemize}
In section~\ref{se:calibration}, we will use the rating and default data for 2009 from table~\ref{tab:all}
as an example for the estimation period (i.e.\ $h =$ 1~year, $T_0 =$ January 1, 2009, $T_0 + h =$ December 31, 2009). 
We will consider both 2010 and 2011 as examples of one-year forecast periods based on the estimation of 
a model for 2009 (i.e.\
$T_1 =$ January 1, 2010 or $T_1 =$ January 1, 2011). The gap of one year between the estimation period 2009 and
the forecast period 2011 reflects the gap that is likely to occur in practice when the full data from the estimation
period usually becomes available only a couple of months after the end of the period.


\section{Description of the model}
\label{se:mecha}

This section describes a statistical model of a borrower's beginning of the
period rating grade and end of the period state of solvency. This model
is applicable to both the estimation and the forecast periods as discussed in section~\ref{se:consequences}.
In particular, we will consider the following model characteristics and
their relationships:
\begin{itemize}
	\item Unconditional rating distribution (profile).
	\item Conditional (on default and survival) rating distributions (profiles).
	\item Unconditional PD.
	\item PD curve (PDs conditional on rating grades).
	\item Accuracy ratio (discriminatory power).
	\item Likelihood ratio.
\end{itemize}
We rely on the standard binary classification model used for topics like pattern recognition, medical diagnoses, or
signal detection \citep[see, e.g.,][]{vanTrees}. \citet{Hand97} presents a variety of
applications (including credit scoring) for this type of model.

Speaking in technical terms, we study the joint distribution and 
some estimation aspects of a pair $(X, S)$ of random variables. The variable $X$ is 
interpreted as the 
\emph{rating grade}\footnote{%
In practice, often a rating model with a small finite number of grades is derived from a
score function with values on a continuous scale. This is usually done by mapping
score intervals on rating grades. See \citet[][section 3]{Tasche2008a} for
a discussion of how such mappings can be defined. Discrete rating models are preferred by
practitioners because manual adjustment of results (overriding) is feasible. Moreover,
by construction results by discrete rating models tend to be more stable over time.}
assigned to a solvent borrower at the beginning of the observation period.
Hence $X$ typically takes on values on  a discrete
scale in a finite set which we describe without loss of
generality as $\{1, 2, \ldots, k\}$. This implies that the marginal distribution
of $X$ is characterised by the probabilities $\Pr[X=x]$, $x=1, \ldots, k$, which
we call the \emph{unconditional rating profile}.

\textbf{Assumption.} \emph{Low values of $X$ indicate low creditworthiness (``bad''), 
high values of $X$ indicate high creditworthiness (``good'').}

The variable $S$ is the \emph{borrower's state of solvency} at the end of  
the observation period, typically one year after the rating grade was observed. 
$S$ takes on values in
$\{0,1\}$. The meaning of $S=0$ is ``borrower has remained solvent'' (solvency or survival), 
$S=1$ means ``borrower has become insolvent'' (default). In particular,
	$S$ is always observed with a time lag to the observation of $X$. 
	Hence, when $S$ is observed $X$ is already known but when $X$ is
	observed today $S$ is still unknown.
We write $D$ for the event $\{S=1\}$ and $N$ for the event $\{S=0\}$.
Hence 
\begin{equation}
	D \cap N = \{S=1\} \cap \{S=0\} = \emptyset, \quad D \cup N = \text{whole space}.
\end{equation}
The marginal distribution of the state variable 
$S$ is characterised by
the \emph{unconditional probability of default} $p$ which is defined as
\begin{equation}\label{eq:PDunconditional}
	p\ =\ \Pr[D]\ =\ \Pr[S =1]\ \in\ [0,1].
\end{equation}
$p$ is sometimes also called \emph{base probability of default}. In the following
we assume $0 < p < 1$ as the cases $p = 0$ and $p=1$ are not
of practical relevance.


\subsection{Model specification}
\label{se:spec}

Recall that the two marginal distributions of $X$ and $S$ respectively
do not uniquely determine the joint distribution of $X$ and $S$. For easy 
reference we state in the following proposition
the three equivalent standard ways to characterise
the joint distribution.
\begin{subequations}
\begin{proposition}\label{pr:joint}
The joint distribution of the pair $(X,S)$ of the rating variable $X$ and
the state of the borrower variable $S$ is fully specified in
any of the following three ways:
\begin{enumerate}
	\item[(i)] By the joint probabilities 
\begin{equation}
\begin{split}
	\Pr[X=x, S=0] & \ =\ \Pr[\{X=x\}\cap N], \quad x = 1, \ldots, k, \quad\text{and}\\
	\Pr[X=x, S=1] & \ =\ \Pr[\{X=x\}\cap D], \quad x = 1, \ldots, k.
\end{split}	
\end{equation}
\item[(ii)] By the unconditional PD $p = \Pr[D] = 1-\Pr[N]$ and
the distributions of $X$ conditional on $D$ and $N$ respectively:
\begin{equation}\label{eq:cond_dist}
	\begin{split}
	 \Pr[X = x\,|\,D] & = \frac{\Pr[\{X = x\}\,\cap D]}{p},\ x = 1,\ldots,k, \quad\text{and}\\
	 \Pr[X = x\,|\,N] & = \frac{\Pr[\{X = x\}\,\cap N]}{1-p}, \ x = 1,\ldots,k. 	 
	\end{split}
\end{equation}
$x\mapsto\Pr[X = x\,|\,D]$ and $x\mapsto\Pr[X = x\,|\,N]$ are called the \emph{conditional
rating profiles} (conditional on default and survival respectively). In a more concise manner
$x\mapsto\Pr[X = x\,|\,D]$ is also called \emph{default (rating) profile} and 
$x\mapsto\Pr[X = x\,|\,N]$ is called \emph{survival (rating) profile}.
\item[(iii)] By the unconditional rating profile $x \mapsto \Pr[X=x]$ and
the \emph{conditional PDs}
\begin{equation}
	\Pr[D\,|\,X=x] \ =\ \frac{\Pr[\{X=x\}\cap D]}{\Pr[X=x]},  \quad x = 1, \ldots, k.
\end{equation}
$x\mapsto\Pr[D\,|\,X=x]$ is called the \emph{PD curve} associated with the grades
$x = 1, \ldots, k$.
\end{enumerate}
\end{proposition}
\end{subequations}

\begin{subequations}
For further reference we note how the specification of the joint distribution of $(X,S)$
given in proposition~\ref{pr:joint}~(ii) implies 
the representation provided in 
proposition~\ref{pr:joint}~(iii): 
\begin{itemize}
	\item By the law of total probability, the unconditional rating profile $\Pr[X = x]$, 
$x = 1,\ldots,k$  can be calculated as
\begin{equation}\label{eq:unconditional}
	\Pr[X = x] \ =\ p\,\Pr[X = x\,|\,D] + (1-p)\,\Pr[X = x\,|\,N].
\end{equation}
	\item Bayes' formula implies the following representation of
the PD curve $\Pr[D\,|\,X=x]$:
\begin{equation}\label{eq:pd_discrete}
		\Pr[D\,|\,X = x] \ =\ \frac{p\,\Pr[X = x\,|\,D]}{p\,\Pr[X = x\,|\,D] +
		(1-p)\,\Pr[X = x\,|\,N]}.
\end{equation}
\end{itemize}
\end{subequations}
\begin{subequations}
Also for further reference, we observe how the specification of the joint distribution of $(X,S)$
given in proposition~\ref{pr:joint}~(iii) implies 
the representation provided in 
proposition~\ref{pr:joint}~(ii): 
\begin{itemize}
	\item Again by the law of total probability, the unconditional PD $p$  can be calculated as
	\begin{equation}\label{eq:uncondPD}
	p \ = \ \sum_{x=1}^k \Pr[D\,|\,X = x]\,\Pr[X = x].
	\end{equation}
	\item With regard to the conditional rating profiles, it follows directly from the 
	definition of conditional probability that
\begin{align}	
	\Pr[X = x\,|\,D]  &\ =\ \Pr[D\,|\,X = x]\,\Pr[X = x] \,/\, p, \quad\text{and}\label{eq:def.profile}\\
\Pr[X = x\,|\,N]  &\ =\ \bigl(1-\Pr[D\,|\,X = x]\bigr)\,\Pr[X = x] \,/\, (1-p).\label{eq:ndef.profile}
\end{align}
\end{itemize}
\end{subequations}
The equivalence between equations \eqref{eq:unconditional} and \eqref{eq:pd_discrete} 
on the one hand and equations \eqref{eq:uncondPD}, \eqref{eq:def.profile} and \eqref{eq:ndef.profile} 
on the other hand allows the calculation of one set of characteristics once the other set of characteristics
is known. But the equivalence also represents a consistency condition that must be kept in mind if one of the
characteristics is changed. In particular, if for a given unconditional rating profile there are
independent estimates of the unconditional PD and the PD curve, equation \eqref{eq:uncondPD} becomes
a crucial consistency condition.

\begin{subequations}
The following proposition presents another consistency condition based on \eqref{eq:unconditional} 
that proves useful in section~\ref{se:calibration} below. We omit its easy proof.
\begin{proposition}\label{pr:consistency}
Let $\pi_x$, $x=1, \ldots, k$ and $q_x$, $x=1, \ldots, k$ be probability distributions and fix
a number $p \in (0,1)$.
\begin{itemize}
	\item[(i)] Define numbers $u_x$, $x=1, \ldots, k$ by solving the following equations for $u_x$:
	\begin{equation}
		\pi_x \ = \ p\,u_x + (1-p)\,q_x, \quad x=1, \ldots, k.
	\end{equation}
	Then $u_x$, $x=1, \ldots, k$ is a proper probability distribution if and only if the following 
	two inequalities hold for all $x=1, \ldots, k$:
	\begin{equation}
		\begin{split}
		p\,q_x &\ \le \pi_x,\\
		(1-p)\,(1-q_x) &\ \le \ 1-\pi_x.
		\end{split}
	\end{equation}
\item[(ii)] Define numbers $v_x$, $x=1, \ldots, k$ by solving the following equations for $v_x$:
	\begin{equation}
		\pi_x \ = \ p\,q_x + (1-p)\,v_x, \quad x=1, \ldots, k.
	\end{equation}
	Then $v_x$, $x=1, \ldots, k$ is a proper probability distribution if and only the following 
	two inequalities hold for all $x=1, \ldots, k$:
	\begin{equation}
		\begin{split}
		p\,(1-q_x) &\ \le 1-\pi_x,\\
		(1-p)\,q_x &\ \le \ \pi_x.
		\end{split}
	\end{equation}	
\end{itemize}
\end{proposition}
\end{subequations}

In this paper, we use quasi moment matching (QMM) as described in appendix~\ref{se:app} to transform
the grade-level empirical default rates into smoothed PD curves. 
As mentioned in section~\ref{se:consequences}, such smoothing of the empirical PD curve is needed in order to
\begin{itemize}
	\item force monotonicity of the PD curve and
	\item force the PDs to be positive.
\end{itemize}	
Matching in this context means fitting a two-parameter
curve to the empirically observed unconditional default rate and discriminatory power. The discriminatory
power is measured as accuracy ratio whose general formula is given in \eqref{eq:AR}. Using the conditional
rating profiles defined by \eqref{eq:cond_dist} the accuracy ratio can also be described by
\begin{equation}\label{eq:AR.emp}
	\mathrm{AR}\ = \ \sum_{x=2}^k \Pr[X = x\,|\,N]\, \Pr[X \le x-1\,|\,D] - 
			\sum_{x=1}^{k-1} \Pr[X = x\,|\,N]\, \Pr[X \ge x+1\,|\,D].
\end{equation}


\subsection{Likelihood ratio}
\label{se:ratio}

The specification of the
model by unconditional rating profile and PD curve (see proposition~\ref{pr:joint} (iii))
may be inappropriate if we want to combine a forecast period profile with an estimation period
PD curve. For according to equations \eqref{eq:unconditional} and \eqref{eq:pd_discrete}
both components depend upon the unconditional PD -- which might be different in the
estimation and forecast periods.  The likelihood ratio is a concept closely related
to the PDs but avoids the issue of dependence on the unconditional PD. The natural logarithm of 
the likelihood ratio is called \emph{weights of evidence} and is an important concept
in credit scoring \citep[see][for a detailed discussion]{thomas2009consumer}. 
 
In the context of credit ratings, it can be reasonably assumed that all components of
the conditional rating profiles $\Pr[X = x\,|\,D]$ and $\Pr[X = x\,|\,N]$, $x = 1, \ldots, k$ are
positive. For otherwise, there would be rating grades with sure predictions of default and
survival -- which is unlikely to happen with real-world rating models.
We can therefore define the \emph{likelihood ratio} $\lambda$ associated with
the rating model:
\begin{equation}\label{eq:lik.ratio}
	\lambda(x) \ = \ \frac{\Pr[X = x\,|\,N]}{\Pr[X = x\,|\,D]}, \quad x = 1, \ldots, k.
\end{equation}
The likelihood ratio $\lambda(x)$ specifies how much more (or less) likely it is
for a survivors's rating grade to come out as $x$ than for a defaulter's rating
grade. Observe that \eqref{eq:pd_discrete} can be rewritten as
\begin{subequations}
\begin{equation}\label{eq:pd.likratio}
		\Pr[D\,|\,X = x] \ =\ \frac{p}{p +
		(1-p)\,\lambda(x)}, \quad x = 1, \ldots, k.
\end{equation}
This is equivalent to an alternative representation of the likelihood ratio:
\begin{equation}\label{eq:alternative}
	\lambda(x) \ = \ \frac{1-\Pr[D\,|\,X = x]}{\Pr[D\,|\,X = x]}\,\frac{p}{1-p}, 
	\quad x = 1, \ldots, k.
\end{equation}
\end{subequations}
By \eqref{eq:alternative}, the likelihood ratio can alternatively be described
as the ratio of the grade $x$ odds of survival and the unconditional odds of
survival.
\begin{subequations}
By \eqref{eq:def.profile}, \eqref{eq:pd.likratio} also implies
\begin{align}
	\Pr[X = x\,|\,D]&\ =\ \frac{\Pr[X = x]}{p +
		(1-p)\,\lambda(x)}, \quad x = 1, \ldots, k, \label{eq:likdefprofile}\\
		\intertext{and, by taking the sum of all $\Pr[X = x\,|\,D]$}	
		1 &\ = \ \sum_{x=1}^k \frac{\Pr[X = x]}{p +
		(1-p)\,\lambda(x)}.\label{eq:likeqn}
\end{align}
\end{subequations}
This observation suggests that the information borne by the likelihood ratio is
very closely related to the information inherent in the PD curve. More specifically, we 
obtain the following characterisation of \eqref{eq:likeqn} which is basically the likelihood ratio
version of \eqref{eq:uncondPD}.
\begin{subequations}
\begin{theorem}\label{th:unique}
Let $\pi_x >0$, $x=1,\ldots, k$ be a probability distribution. Assume that $x \mapsto \lambda(x)$
is positive for $x=1,\ldots, k$. Consider the equation
\begin{equation}\label{eq:eqn}
	 \sum_{x=1}^k \frac{\pi_x}{p + (1-p)\,\lambda(x)} \ = \ 1.
\end{equation}
Then with regard to solutions $p\in [0,1]$ of \eqref{eq:eqn} other than $p=1$ the 
following statements hold:
\begin{enumerate}
	\item[(i)] Assume that $x \mapsto \lambda(x)$ is a mapping onto a constant, i.e.\ $\lambda(x) = \lambda$ 
	for all $x = 1, \ldots, k$.
	Then all $p\in [0,1]$ are solutions of \eqref{eq:eqn} if $\lambda = 1$ and there is no solution 
	 $p\in [0,1)$ if $\lambda \not= 1$.
	\item[(ii)] Assume that $x \mapsto \lambda(x)$ is \textbf{not} a mapping onto a constant. Then
	there exists a solution $p\in [0,1)$ of \eqref{eq:eqn} if and only if
	\begin{equation}\label{eq:condition}
	\sum_{x=1}^k \frac{\pi_x}{\lambda(x)}\ \ge\ 1 \quad\text{and}\quad \sum_{x=1}^k \pi_x\,\lambda(x)\ >\ 1.
\end{equation}
	If there exists a solution $p\in [0,1)$ of \eqref{eq:eqn} then this solution is unique. The unique
	solution is $p = 0$ if and only if 
\begin{equation}\label{eq:iff}
	\sum_{x=1}^k \frac{\pi_x}{\lambda(x)} \ = \ 1.
\end{equation}
	\end{enumerate}
\end{theorem}
\end{subequations}
\begin{figure}[t!p]
\caption{Illustration for the proof of theorem~\ref{th:unique}. The three
possibilities for the shape of the graph of the function defined by \eqref{eq:f}.}
\label{fig:1}
\begin{center}
\ifpdf
	\includegraphics[width=15cm]{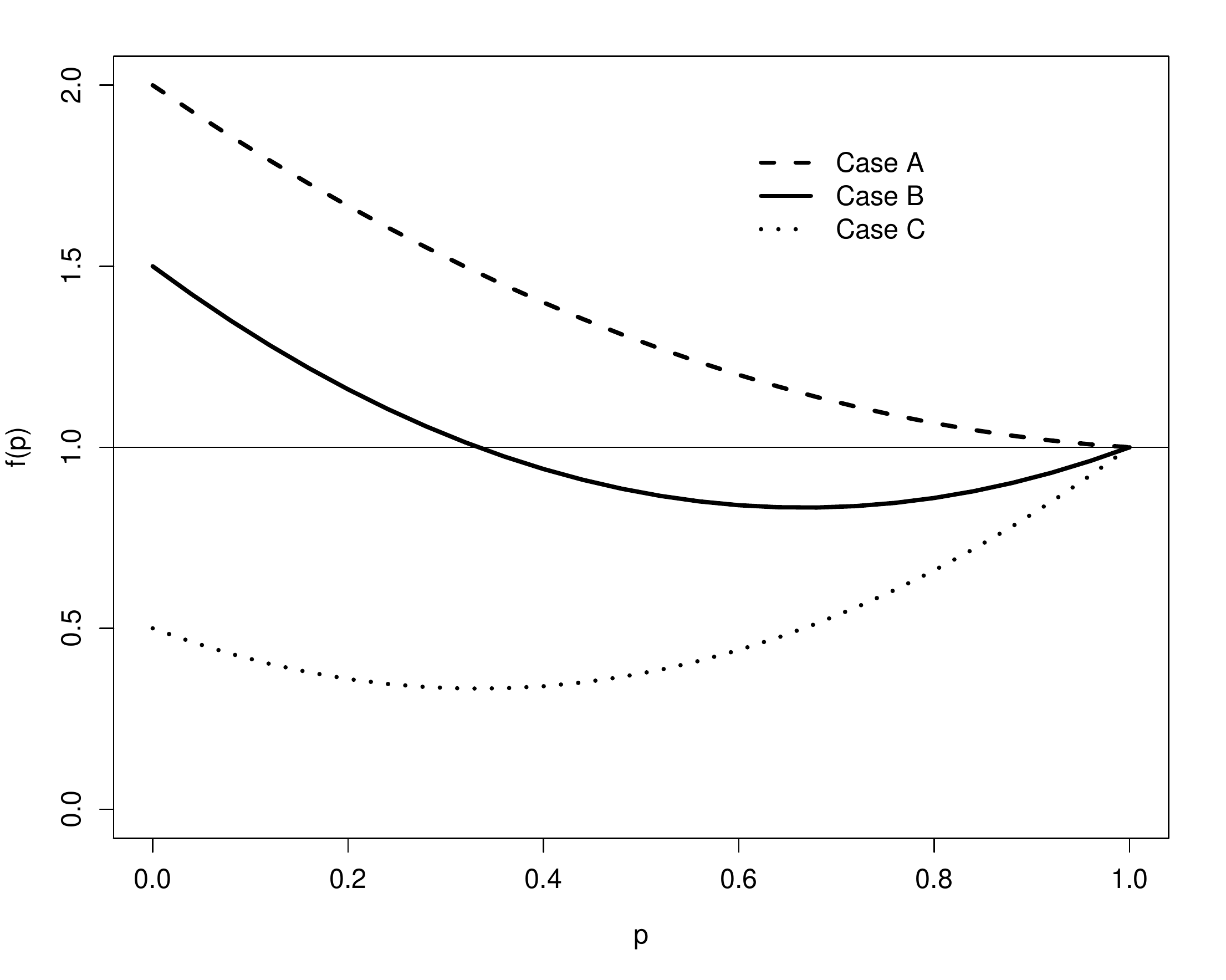}
\fi
\end{center}
\end{figure}
\textbf{Proof.} Statement (i) is obvious. With regard to statement (ii), define the function
$f: [0,1] \to (0,\infty)$ by
\begin{subequations}
\begin{equation}\label{eq:f}
	f(p) \ = \ \sum_{x=1}^k \frac{\pi_x}{p + (1-p)\,\lambda(x)}.
\end{equation}
Observe that $f$ is twice continuously differentiable in $p$ with
\begin{align}
	f'(p) & \ = \ \sum_{x=1}^k \frac{(\lambda(x)-1)\,\pi_x}{(p + (1-p)\,\lambda(x))^2}
	\quad\text{and}\label{eq:fprime}\\
	f''(p) & \ = \ 2 \sum_{x=1}^k \frac{(\lambda(x)-1)^2\,\pi_x}{(p + (1-p)\,
	\lambda(x))^3}.\label{eq:fprimeprime}
\end{align}
From \eqref{eq:f} and \eqref{eq:fprime} we obtain
\begin{equation}\label{eq:values}
	f(0) \ =\ \sum_{x=1}^k \frac{\pi_x}{\lambda(x)},\quad f(1) \ =\ 1,\quad\text{and}\quad
	f'(1) \ =\ \sum_{x=1}^k \pi_x\,\lambda(x) - 1.
\end{equation}
\end{subequations}
\eqref{eq:fprimeprime} implies $f''(p) > 0$ because $\lambda(x)$ is not constant
by assumption. Hence $f$ is strictly convex in $p$.
The strict convexity of $f$ implies that the shape of the graph of $f$ is determined
by \eqref{eq:values} and that only the following three cases can occur:
\begin{eqnarray*}
\text{case A:}    \quad & f(0)\ > \ 1\quad\text{and}\quad f'(1) \ \le\ 0, \\
\text{case B:}  \quad & f(0)\ \ge \ 1\quad\text{and}\quad f'(1) \ >\ 0, & \quad\text{or}\\
\text{case C:}  \quad & f(0)\ < \ 1\quad\text{and}\quad f'(1) \ >\ 0. &
\end{eqnarray*}
A stylised illustration of the three different possible shapes of the graph of
$f$ is shown in figure~\ref{fig:1} on page~\pageref{fig:1}. Only in case~B there is a second (and only one) intersection at
a $p <1$ of the horizontal line through~1. By \eqref{eq:values}, case~B is equivalently
described by \eqref{eq:condition}. The second intersection of the horizontal line through~1 occurs
at $p=0$ if and only if $f(0) = 1$ which is equivalent to \eqref{eq:iff}.
\hfill\textbf{q.e.d.}

At first glance, theorem~\ref{th:unique} might appear as an unnessecarily complicated 
way to describe the interplay of unconditional rating profile, likelihood ratio, and unconditional
PD. However, theorem~\ref{th:unique} becomes interesting when we try to construct the joint
distribution of a borrower's rating $X$ at the beginning of the observation period and 
the borrower's state $S$ at the end of the period from an unconditional rating profile
and a candidate likelihood ratio (which might have been estimated separately). In this
context, theorem~\ref{th:unique} tells us that the construction will work only
if condition \eqref{eq:condition} is satisfied. In contrast, by proposition~\ref{pr:joint}~(iii)
the construction is always possible if one combines an unconditional rating profile 
with a candidate PD curve (assuming that all its components take values between 0 and 1).  

Actually, from theorem~\ref{th:unique} it is not yet clear that it gives indeed rise to
a fully specified joint distribution of rating $X$ and default or survival state $S$. This,
however, is confirmed by the next proposition whose straight-forward proof is omitted.

\begin{proposition}\label{pr:full}
Let $\pi_x >0$, $x=1,\ldots, k$ be a probability distribution. Assume that $x \mapsto \lambda(x)$
is positive for $x=1,\ldots, k$ and that equation \eqref{eq:eqn} has a 
solution $0 < p < 1$. Then there exists a unique joint distribution of $X$ and $S$ such
that $x \mapsto \lambda(x)$ is the likelihood ratio associated with the joint distribution
in the sense of equation \eqref{eq:lik.ratio}.
\end{proposition}


\subsection{Smoothing observed default rates}
\label{se:example.estimation}
		
In this section, we illustrate the concepts introduced in sections~\ref{se:spec} and \ref{se:ratio}
by revisiting the S\&P data for 2009 presented in section~\ref{se:data}. As the notation introduced
at the beginning of the section requires we map the S\&P rating symbols CCC-C, B-, B, $\ldots$, AA+, AAA onto
the numbers $1, \ldots, 17$ (hence grade 17 stands for the least risky grade AAA).

\begin{table}[b!p]
\caption{Unconditional default rate and accuracy ratio
for the 2009  
corporate data from table~\ref{tab:all}.}
\label{tab:moments}
\begin{center}
\begin{tabular}{|c|c|}
\hline
Default rate & Accuracy ratio  \\ \hline
3.99\% & 82.7\%  \\ \hline
\end{tabular}
\end{center}
\end{table}

\begin{table}[t!p]
\caption{Grade-level default rates and smoothed conditional PDs (PD curve)
for the 2009 corporate  
corporate data from table~\ref{tab:all}. All
numbers in \%.}
\label{tab:pd.curves2009}
\begin{center}
\begin{tabular}{|l||c|c|}
\hline
Rating grade & Default rate & Smoothed PD \\ \hline \hline
AAA & 0.000 & 0.003 \\ \hline
AA+ & 0.000 & 0.006 \\ \hline
AA & 0.000 & 0.012 \\ \hline
AA- & 0.000 & 0.025 \\ \hline
A+ & 0.294 & 0.047 \\ \hline
A & 0.392 & 0.091 \\ \hline
A- & 0.000 & 0.173 \\ \hline
BBB+ & 0.402 & 0.299 \\ \hline
BBB & 0.185 & 0.495 \\ \hline
BBB- & 1.089 & 0.797 \\ \hline
BB+ & 0.000 & 1.138 \\ \hline
BB & 1.017 & 1.518 \\ \hline
BB- & 0.907 & 2.280 \\ \hline
B+ & 5.479 & 3.943 \\ \hline
B & 9.959 & 7.999 \\ \hline
B- & 17.162 & 19.557 \\ \hline
CCC-C & 48.421 & 48.355 \\ \hline
\end{tabular}
\end{center}
\end{table}

Column 2 of table~\ref{tab:profiles} shows the unconditional profile
	$x \mapsto \Pr[X=x]$, $\{1, \ldots, 17\} \to [0,1]$ for the S\&P corporate ratings.

Table \ref{tab:moments} on page~\pageref{tab:moments} shows the empirical unconditional default rate and accuracy ratio
	for our estimation data (i.e.\ the 2009 S\&P data). The accuracy ratio
	was calculated according to \eqref{eq:AR.emp}.
	
	Table \ref{tab:pd.curves2009} on page~\pageref{tab:pd.curves2009} presents both the empirically observed grade-level 
	default rates and the smoothed PD curve (according to appendix~\ref{se:app}, with the values
	from table~\ref{tab:moments} as targets) for 
	the 2009 S\&P data. It is hard to assess directly from the numbers 
	how well or badly the smoothed curves fit the empirical data. Therefore
	we calculate an \emph{implied default profile} and compare it by means of
	a $\chi^2$ test with the observed default profile. `Implied default profile' means
	the theoretical rating distribution conditional on default that is derived by means of 
	Equation~\eqref{eq:def.profile} from the unconditional rating profile, the PD curve and the unconditional PD.

Table \ref{tab:implied.def.profiles2009} on page~\pageref{tab:implied.def.profiles2009} shows the empirically
	observed and implied default profiles for the 2009 S\&P data. Clearly the fit is not perfect. This impression is confirmed by application of the 
	$\chi^2$
	test mentioned in section~\ref{se:observations}. To apply the test, choose the implied profile as
	Null hypothesis distribution and test against the default numbers for 2009 as given in table~\ref{tab:all}. 
	
	The test result 
	(based on Monte-Carlo approximation) is
	a p-value of 9\% for the test of the fit of the implied corporates default profile. 
	Hence the fit could be rejected as too poor at 10\% type-I error level.
	However, given the inversions of default rates in the corporates data it
	might be hard to get a much better fit with any other forced monotonic PD curve
	estimate. We therefore adopt the corporates  smoothed
	PD curve from table~\ref{tab:pd.curves2009} as a starting point for the PD curve calibration
	examples described in section~\ref{se:calibration} below.
\begin{table}[t!p]
\caption{Empirical and implied default profiles
for the 2009 
corporate data from table~\ref{tab:all}. All
numbers in \%.}
\label{tab:implied.def.profiles2009}
\begin{center}
\begin{tabular}{|l||c|c|}
\hline
Rating grade & Empirical profile & Implied profile \\ \hline \hline
AAA & 0.0000 & 0.0010 \\ \hline
AA+ & 0.0000 & 0.0009 \\ \hline
AA & 0.0000 & 0.0094 \\ \hline
AA- & 0.0000 & 0.0261 \\ \hline
A+ & 0.4274 & 0.0685 \\ \hline
A & 0.8547 & 0.1989 \\ \hline
A- & 0.0000 & 0.4041 \\ \hline
BBB+ & 0.8547 & 0.6355 \\ \hline
BBB & 0.4274 & 1.1436 \\ \hline
BBB- & 2.1368 & 1.5642 \\ \hline
BB+ & 0.0000 & 1.2934 \\ \hline
BB & 1.2821 & 1.9143 \\ \hline
BB- & 1.7094 & 4.2968 \\ \hline
B+ & 10.2564 & 7.3802 \\ \hline
B & 20.5128 & 16.4769 \\ \hline
B- & 22.2222 & 25.3242 \\ \hline
CCC-C & 39.3162 & 39.2628 \\ \hline\hline
All & 100.0000 & 100.0000 \\ \hline
\end{tabular}
\end{center}
\end{table}	


\section{Calibration approaches}
\label{se:calibration}

The result of the estimation period is a fully specified (and smoothed) model
for the joint distribution of a borrower's beginning of
the period rating $X$ and  end of the period solvency state $S$. 
In this section, we discuss how to combine the estimation period model
with observations from the beginning of the forecast period 
in order to predict the grade-level default rates that are observed
at the end of the forecast period. This process is often referred to 
as \emph{calibration of the PD curve.}

\textbf{Notation.} \emph{All objects (like probabilities and the likelihood ratio) from the 
estimation period are labelled with subscript 0. All objects 
from the forecast period are labelled with subscript 1.}

In the following we will make use, in particular, of assumptions
on the invariance or specific transformation between estimation and 
forecast period of
\begin{itemize}
	\item the conditional rating profiles $\Pr_0[X=x\,|\,D]$ and
	$\Pr_0[X=x\,|\,N]$ (for $x=1, \ldots, k$),
	\item the PD curve $x \mapsto \Pr_0[D\,|\,X = x]$, and
	\item the likelihood ratio $x \mapsto \lambda_0(x)$.
\end{itemize}

Imagine we are now at the beginning of the forecast period. The borrowers'
states of solvency at the end of the period are yet unknown. 
The objective of the forecast period is to predict the default
rates to be observed at the end of the period for the rating grades $1, \ldots, k$
by conditional PDs (PD curve) $\Pr_1[D\,|\,X=x]$, $x=1, \ldots, k$. There are
different forecast approaches for the conditional PDs. The selection
of a suitable approach, in particular, depends on what we already
know at the beginning of the forecast period about the joint distribution of 
a borrower's rating $X$ at the beginning of
the period and the borrower's solvency state $S$ at the end of the
period. We will look in detail at the following two
possibilities:
\begin{itemize}
	\item \emph{The unconditional rating profile $\Pr_1[X=x]$, $x=1, \ldots, k$
	is known}. This is likely to be the case for a newly developed rating model
	if all borrowers can be re-rated with the new model in a big-bang effort 
	before the beginning of the forecast period. It will also be the case
	if an existing rating model is re-calibrated. Even if in the case of
	a new rating model no timely re-rating of the whole portfolio is feasible, it
	might still be possible (and should be tried) to re-rate a representative 
	sample of the borrowers in the portfolio such that a reliable estimate of the
	unconditional rating profile is available. Where this is not
	possible, the
	rating model should be used in parallel run with the incumbent rating
	model until such time as the full rating profile of the portfolio
	has been determined. Only then a PD curve forecast with some chance
	of being accurate can be made. This might be one of the reasons
	for the  `credible track record' requirement of the Basel Committee
	\citep[][paragraph~445]{BaselAccord}. However, we will see in 
	section~\ref{se:case3} that as soon as a forecast of
	the unconditional PD is given a meaningful if not accurate
	PD curve forecast can be made without knowledge of the actual
	unconditional rating profile. This forecast could be used
	for a preliminary calibration during the Basel~II `track record' period.
	\item \emph{An estimate of the unconditional PD $p_1$ for the
	forecast period is available.} This forecast could be a proper best estimate, a pessimistic estimate
	for stress testing purposes, or a long-run estimate for the purpose
	of a through-the-cycle (TTC) calibration\footnote{%
See \citet{Heitfield2005} for a detailed discussion of point-in-time (PIT) and TTC
rating and PD estimation approaches. See \citet{loffler2012} for the question
of how much TTC agency ratings are.}. 
\end{itemize}
These two possibilities are not exclusive nor do they necessarily occur together. That
is why, in the following, we discuss four cases:
\begin{itemize}
	\item \textbf{Case 1.} The unconditional rating profile for the forecast period is known and an 
	independent estimate of the unconditional PD is available.
	\item \textbf{Case 2.} The unconditional rating profile for the forecast period is not known but an 
	independent estimate of the unconditional PD is available.
	\item \textbf{Case 3.} The unconditional rating profile for the forecast period is known but no 
	independent estimate of the unconditional PD is available.
	\item \textbf{Case 4.} Neither the unconditional rating profile nor the unconditional PD for the
	forecast period are known.
\end{itemize}
For each of the four cases we will present one or more approaches to estimate a set of
model components needed to specify a full model. See proposition~\ref{pr:joint} 
for the main possibilities to specify a full proper model of a
borrower's beginning of the period rating and end of the period solvency state.
We will illustrate the forecast approaches presented with numerical examples
based on the S\&P data from table~\ref{tab:all}.
In none of the four cases there is sufficient information from the forecast period available
to completely specify a model. That is why assumptions about inter-period
invariance of model components play an important role in the forecast process. 

\subsection{Invariance assumptions}
\label{se:invariance}

Forecasting without assuming that some of the features observed in the estimation period
are invariant (i.e.\ unchanged) between the estimation and forecast periods is impossible.
Ideally, any assumption of invariance should be theoretically sound, and it should be
possible to verify it by backtesting. In this section, we briefly discuss which
invariance assumptions for the model from section~\ref{se:mecha} 
we should look at closer in the following.
\begin{itemize}
	\item It is obvious that no invariance assumptions must be made on
	objects that can be observed or reliably estimated in a separate 
	forecast exercise at the beginning of the forecast period: 
\begin{itemize}
	\item As explained above, in particular, 
	the actual unconditional rating profile of the portfolio should be
	known at the beginning of the observation period. 
	\item We look both at the case that the forecast unconditional
	default rate is estimated based on the unconditional forecast period
	rating profile and at the case where an independent forecast of the forecast
	period unconditional default rate is available.
\end{itemize}
	\item As the future solvency states of the borrowers in the portfolio
	are not yet known at the beginning of the forecast period, assuming
	that both conditional rating profiles are invariant could make sense. 
	\item Assuming that the likelihood ratio is invariant is less restrictive than
	the assumption of invariant conditional rating profiles. 
	\item Instead of assuming that both conditional rating profiles are invariant, one
	could also assume that only one of the two is invariant. If we assume that the
	survival profile $x\mapsto\Pr[X = x\,|\,N]$ is invariant then proposition~\ref{pr:consistency} 
	implies for all rating grades $x$ the restriction
	\begin{equation}\label{eq:restrict}
		(1-p_1)\,\Pr_0[X = x\,|\,N] \ \le \ \Pr_1[X = x].
	\end{equation}
	Probabilities of default are often measured at a one year horizon. In that 
	case the forecast $p_1$ of the unconditional PD for principal portfolios like banks
	or corporates will hardly ever exceed 5\%. This implies, however, that condition
	\eqref{eq:restrict} is easily violated. In practice, therefore, quite often
	the assumption of an invariant survival rating profile will not result in a proper model.
	That is why we do not discuss further details of this invariance assumption in this 
	paper.
	\item Assuming the default rating profile as invariant is a much more promising approach because 
	the conditions for the default profile to generate a proper model are much easier satisfied than
	condition \eqref{eq:restrict} for the survival profile.
	\item Invariance assumptions may be weakened by restating them as shape invariance
	assumptions.
\begin{itemize}
	\item For instance, a common approach is to assume that the shape of the PD curve 
	is preserved between the estimation and the forecast periods. This can be accomplished
	by scaling the PD curve with a constant multiplier that is determined at the beginning
	of the forecast period \citep[see, e.g.,][page~67]{Falkenstein&Boral&Carty}.
	\eqref{eq:pd_discrete} shows
	that the scaled PD curve strongly depends on the estimation period unconditional PD.
	Hence making use of the scaled PD curve for forecasts in the forecast period might
	`contaminate' the forecast with the estimation period unconditional PD which might be quite different from the 
	forecast period unconditional PD. We include the scaled PDs approach nonetheless in
	the subsequent more detailed discussion because of its simplicity and popularity.
	\item Scaling the likelihood ratio instead of the PD curve avoids the contamination
	issue we have observed for the scaled PD curve. 
\end{itemize}
\end{itemize}
	

\subsection{Case 1: Unconditional rating profile and unconditional PD given}
\label{se:case4}

In this case, it is assumed that the unconditional rating profile $\Pr_1[X=x]$, $x=1, \ldots, k$ 
can directly be observed at the beginning of the forecast period, and it is also assumed 
that a forecast unconditional PD $0 < p_1 < 1$ is given 
that is likely to differ from the estimation period unconditional PD.
 There are several approaches to prediction in the forecast period that
may lead to proper models for the forecast period:
\begin{itemize}
	\item \textbf{Invariant default profile.} Assume that the default rating profile is invariant, i.e.
\begin{equation}\label{eq:defprofile.invariant}
		\Pr_1[X=x\,|\,D]  \ = \ \Pr_0[X=x\,|\,D], \quad x=1, \ldots, k.	
\end{equation}
	\item \textbf{Invariant AR.} Assume that the discriminatory power of the model as measured
	by the accuracy ratio (see \eqref{eq:AR.emp}) is invariant, i.e.
\begin{equation}\label{eq:AR.invariant}
		\mathrm{AR}_1  \ = \ \mathrm{AR}_0.	
\end{equation}
	\item \textbf{Scaled PDs.} Assume that the estimation period PD curve can be linearly scaled
	to become the forecast period PD curve, i.e.\ there is a constant $c_\mathrm{PD} > 0$
	such that
\begin{equation}\label{eq:PDcurve.scaled}
	\Pr_1[D\,|\,X = x] \ =\ c_\mathrm{PD}\,\Pr_0[D\,|\,X = x], \quad x=1, \ldots, k.
\end{equation}
	\item \textbf{Scaled likelihood ratio.} Assume that the estimation period likelihood ratio can be linearly scaled
	to become the forecast period likelihood ratio, i.e.\ there is a constant $c_\mathrm{LR} > 0$
	such that
\begin{equation}\label{eq:likratio.scaled}
	\lambda_1(x) \ =\ c_\mathrm{LR}\,\lambda_0(x), \quad x=1, \ldots, k.
\end{equation}
\end{itemize}	
In principle, a fifth approach is cogitable, namely to assume that the survivor
rating profile does not change from the estimation period to the forecast period. However,
as explained in section~\ref{se:invariance} it is unlikely that this approach results in a proper forecast period
model with a proper default rating profile. That is why we do not discuss this
approach.
\begin{subequations}
\subsubsection{On assumption \eqref{eq:defprofile.invariant}} This assumption is not necessarily
viable as \eqref{eq:unconditional} must be satisfied. It follows from proposition~\ref{pr:consistency} that assumption
\eqref{eq:defprofile.invariant} makes for a proper model of a borrower's rating 
and state of solvency if and only if we have for all $x=1, \ldots, k$
\begin{equation}\label{eq:proper}
	p_1\,\Pr_0[X=x\,|\,D] \: \le\: \Pr_1[X=x]\quad \text{and}\quad
	p_1\,\bigl(1-\Pr_0[X=x\,|\,D]\bigr) \: \le\: \bigl(1-\Pr_1[X=x]\bigr).
\end{equation}
If \eqref{eq:proper} holds then by \eqref{eq:def.profile} we obtain the
following equation for the PD curve:
\begin{equation}
	\Pr_1[D\,|\,X = x] \ =\ \frac{p_1\,\Pr_0[X = x\,|\,D]}{\Pr_1[X = x]}.
\end{equation}
\end{subequations}
Actually, there are two slightly different approaches to implement assumption~\eqref{eq:defprofile.invariant}:
\begin{itemize}
	\item[(i)] Use a smoothed version of the estimation period default profile that could be derived 
	via equation~\eqref{eq:def.profile} from a smoothed PD curve -- which in turn might have been
	determined by QMM as described in appendix~\ref{se:app}.
	\item[(ii)] Use the observed estimation period default profile and the given forecast period unconditional profile 
	to determine by means of \eqref{eq:ndef.profile} an implied raw survivor profile. Based on this survivor
	profile and the observed estimation period default profile deploy
	equation~\eqref{eq:AR.emp} to compute a forecast accuracy ratio. Apply then QMM as described in appendix~\ref{se:app}
	to determine a smoothed PD curve for the forecast period.
\end{itemize}
Compared with approach (i), approach (ii) has the advantage of always delivering a monotonic PD curve. That is why
for the purpose of this paper we implement assumption~\eqref{eq:defprofile.invariant} in the shape of (ii) 
although anecdotal evidence shows that the performance of (ii) is not necessarily better than the performance of (i).

\begin{table}[t!p]
\caption{2010 and 2011 grade-level forecast default rates for
S\&P corporates ratings. P-values
are for the $\chi^2$-tests of the implied default profiles.
All values in \%.}
\label{tab:All.PDcurves}
\begin{center}
\begin{tabular}{|l||c|c|c|c|c|}
\hline
& Default rate & Invariant default & Invariant AR & Scaled PDs & Scaled likelihood \\ 
&  & profile \eqref{eq:defprofile.invariant} & \eqref{eq:AR.invariant} & \eqref{eq:PDcurve.scaled} & 
ratio  \eqref{eq:likratio.scaled} \\ \hline\hline 
& \multicolumn{5}{c|}{2010: Unconditional default rate 1.141} \\ \hline\hline
AAA & 0 & 0.0012 & 0.0004 & 0.0007 & 0.0005 \\ \hline
AA+ & 0 & 0.0023 & 0.0009 & 0.0015 & 0.0012 \\ \hline
AA & 0 & 0.0041 & 0.0018 & 0.0031 & 0.0023 \\ \hline
AA- & 0 & 0.0083 & 0.0040 & 0.0066 & 0.0049 \\ \hline
A+ & 0 & 0.0163 & 0.0086 & 0.0125 & 0.0093 \\ \hline
A & 0 & 0.0319 & 0.0183 & 0.0241 & 0.0180 \\ \hline
A- & 0 & 0.0593 & 0.0366 & 0.0458 & 0.0342 \\ \hline
BBB+ & 0 & 0.0995 & 0.0652 & 0.0789 & 0.0590 \\ \hline
BBB & 0 & 0.1647 & 0.1145 & 0.1307 & 0.0979 \\ \hline
BBB- & 0 & 0.2660 & 0.1955 & 0.2107 & 0.1581 \\ \hline
BB+ & 0.7874 & 0.3706 & 0.2827 & 0.3006 & 0.2263 \\ \hline
BB & 0.3623 & 0.4847 & 0.3806 & 0.4012 & 0.3029 \\ \hline
BB- & 0.5277 & 0.6907 & 0.5631 & 0.6024 & 0.4576 \\ \hline
B+ & 0.0000 & 1.1043 & 0.9460 & 1.0417 & 0.8023 \\ \hline
B & 0.6881 & 2.0554 & 1.8843 & 2.1134 & 1.6844 \\ \hline
B- & 2.0690 & 4.5380 & 4.5164 & 5.1671 & 4.5716 \\ \hline
CCC-C & 22.2727 & 12.9712 & 14.5179 & 12.7755 & 15.5760 \\ \hline\hline
P-value & Exact & 4.6 & 8.0 & 4.0 & 11.3 \\ \hline\hline 
& \multicolumn{5}{c|}{2011: Unconditional default rate 0.752}  \\ \hline \hline
AAA & 0 & 0.0006 & 0.0003 & 0.0006 & 0.0004 \\ \hline
AA+ & 0 & 0.0013 & 0.0006 & 0.0012 & 0.0009 \\ \hline
AA & 0 & 0.0024 & 0.0013 & 0.0024 & 0.0018 \\ \hline
AA- & 0 & 0.0048 & 0.0027 & 0.0050 & 0.0039 \\ \hline
A+ & 0 & 0.0095 & 0.0058 & 0.0095 & 0.0074 \\ \hline
A & 0 & 0.0186 & 0.0120 & 0.0183 & 0.0143 \\ \hline
A- & 0 & 0.0345 & 0.0236 & 0.0347 & 0.0271 \\ \hline
BBB+ & 0 & 0.0579 & 0.0416 & 0.0599 & 0.0468 \\ \hline
BBB & 0 & 0.0923 & 0.0691 & 0.0992 & 0.0777 \\ \hline
BBB- & 0.1969 & 0.1468 & 0.1147 & 0.1600 & 0.1256 \\ \hline
BB+ & 0 & 0.2065 & 0.1662 & 0.2282 & 0.1797 \\ \hline
BB & 0 & 0.2694 & 0.2219 & 0.3046 & 0.2405 \\ \hline
BB- & 0 & 0.3828 & 0.3248 & 0.4573 & 0.3635 \\ \hline
B+ & 0.3929 & 0.6291 & 0.5567 & 0.7909 & 0.6378 \\ \hline
B & 1.1945 & 1.3483 & 1.2710 & 1.6046 & 1.3414 \\ \hline
B- & 3.9867 & 3.7460 & 3.7941 & 3.9231 & 3.6627 \\ \hline
CCC-C & 15.9420 & 12.1942 & 13.3871 & 9.6998 & 12.7721 \\ \hline\hline
P-value & Exact & 78.5 & 89.4 & 36.6 & 82.2 \\ \hline
\end{tabular}
\end{center}
\end{table}

\subsubsection{On assumption \eqref{eq:AR.invariant}} Actually, even with unconditional
rating profile, unconditional PD, and accuracy ratio given the joint distribution
of a borrower's beginning of the period rating and end of the period state is not uniquely
determined. We suggest applying QMM as in the estimation period 
(see section~\ref{se:example.estimation}) and described in appendix~\ref{se:app}
to compute a PD curve as a forecast of the grade-level default rates. There
is, however, the problem that QMM requires the rating profile conditional on
survival as an input -- which cannot be observed or implied at this stage.
But QMM is fairly robust with regard to the frequencies of the rating grades used
as input to the algorithm. That is why approximating the rating profile conditional on
survival with the unconditional rating profile (known by assumption) seems to work reasonably well.

\begin{subequations}
\subsubsection{On assumption \eqref{eq:PDcurve.scaled}} The constant $c_{\mathrm{PD}}$ is determined
by equation~\eqref{eq:uncondPD}:
\begin{equation}
	c_{\mathrm{PD}} \ =\ \frac{p_1}{\sum_{x=1}^k \Pr_0[D\,|\,X = x]\,\Pr_1[X = x]}
\end{equation}
However, if $c_{\mathrm{PD}} > 1$ the resulting model could be improper because
by \eqref{eq:PDcurve.scaled} it could turn out that $\Pr_1[D\,|\,X = x] > 1$ for
some $x$. If the resulting model under assumption \eqref{eq:PDcurve.scaled}
is proper the implied default profile is as follows:
\begin{equation}
	\Pr_1[X = x\,|\,D]  \ =\ c_{\mathrm{PD}}\,\Pr_0[D\,|\,X = x]\,\Pr_1[X = x]\, /\, p_1,
	\quad x=1, \ldots, k.
\end{equation}
\end{subequations}

\subsubsection{On assumption \eqref{eq:likratio.scaled}} By \eqref{eq:likeqn} we 
obtain an equation that determines the constant $c_\mathrm{LR}$:
\begin{subequations}
\begin{equation}\label{eq:lik.constant}
	1 \ = \ \sum_{x=1}^k \frac{\Pr_1[X = x]}{p_1 + (1-p_1)\,c_\mathrm{LR}\,\lambda_0(x)}.
\end{equation}
Note that
\begin{equation*}
	\lim\limits_{c\to\infty} \sum_{x=1}^k \frac{\Pr_1[X = x]}{p_1 + (1-p_1)\,c\,\lambda_0(x)} = 0\quad
	\text{and}\quad \lim\limits_{c\to 0} \sum_{x=1}^k \frac{\Pr_1[X = x]}{p_1 + (1-p_1)\,c\,\lambda_0(x)} = 1/p_1 > 1.
\end{equation*}
Hence equation \eqref{eq:lik.constant} has always a unique solution $c_\mathrm{LR} > 0$. By
proposition~\ref{pr:full} then we know that under assumption~\eqref{eq:likratio.scaled}
we have a proper model of a borrower's rating and default state. In addition,
by theorem~\ref{th:unique} the resulting forecast likelihood ratio 
$\lambda_1(x)=c_\mathrm{LR}\,\lambda_0(x)$ satisfies the
inequalities
\begin{equation*}
	\sum_{x=1}^k \frac{\Pr_1[X = x]}{\lambda_1(x)}\ >\ 1 \quad\text{and}
	\quad \sum_{x=1}^k \Pr_1[X = x]\,\lambda_1(x)\ >\ 1.
\end{equation*}
This implies the following inequalities for $c_\mathrm{LR}$:
\begin{equation}\label{eq:useful}
	\frac 1{\sum_{x=1}^k \Pr_1[X = x]\,\lambda_0(x)} \ <\ c_\mathrm{LR} \ < \ 
	\sum_{x=1}^k \frac{\Pr_1[X = x]}{\lambda_0(x)}.
\end{equation}
\eqref{eq:useful} is useful because it provides initial values for the numerical
solution of \eqref{eq:lik.constant} for $c_\mathrm{LR}$.
Once $c_\mathrm{LR}$ has been determined \eqref{eq:likdefprofile} and 
\eqref{eq:pd.likratio} imply the following equations for
the default profile and the PD curve under assumption \eqref{eq:likratio.scaled}:
\begin{align}
	\Pr_1[X = x\,|\,D] & \ =\ \frac{\Pr_1[X = x]}{p_1 +
		(1-p_1)\,c_\mathrm{LR}\,\lambda_0(x)}, \quad x = 1, \ldots, k,\\[1ex]
	\Pr_1[D\,|\,X = x] & \ =\ \frac{p_1}{p_1 +
		(1-p_1)\,c_\mathrm{LR}\,\lambda_0(x)}, \quad x = 1, \ldots, k.	
\end{align}
\end{subequations}

\subsubsection{Summary of section~\ref{se:case4}} Table \ref{tab:All.PDcurves} on page~\pageref{tab:All.PDcurves}
shows the results of an application of the approaches presented above 
to forecasting the 2010 and 2011 grade-level default rates of the S\&P corporates
portfolio, based on estimates made with
data from 2009. To allow for a fair performance comparison, we have made use of
prophetic estimates of the 2010 and 2011 unconditional default rates, by setting 
the value of $p_1$ to the observed unconditional default rate of the respective
year and sample.

In order to express the performance of the different approaches in one number
for each approach, we have used the forecast PD curves to derive 
forecast default profiles by means of \eqref{eq:def.profile}. 
The forecast default profiles can be $\chi^2$ tested
against the observed grade-level default numbers from table~\ref{tab:all}.
The p-values of these tests are shown in the last rows of the panels of 
table~\ref{tab:All.PDcurves}. Recall that
higher p-values mean better goodness of fit. 

Table \ref{tab:All.PDcurves} hence indicates that under the
constraints of this section (unconditional rating profile and default rate are given)
the scaled likelihood ratio approach~\eqref{eq:likratio.scaled} and the invariant accuracy ratio
approach~\eqref{eq:AR.invariant} work best, followed
by the invariant default profile approach~\eqref{eq:defprofile.invariant}. This anecdotal evidence,
however, does not allow an unconditional conclusion that `scaled likelihood ratio' or 
`invariant accuracy ratio'
are the best approaches to PD curve calibration. We will test this conclusion on
a larger dataset in section~\ref{se:backtest}.

But also from a conceptual angle there might be good reasons to
prefer the `invariant default profile' approach. When a new rating model is developed one has often to combine data from
several observation periods in order to create a sufficiently large training sample.
Estimating the likelihood ratio from such a combined sample would implicitly be based
on the assumption of an invariant likelihood ratio. Hence it would be strange to
modify the likelihood ratio via scaling in the forecast period. This consistency
issue is obviously avoided with the `invariant default profile' and the `invariant
accuracy ratio' approaches. As we have seen, to implement the `invariant
accuracy ratio' approach we need to approximate the forecast period survivor profile
by the forecast period unconditional rating profile. This approximation could be poor
if the forecast period unconditional default rate is high. Hence, depending
on what approach had been followed in the estimation period and how big the 
forecast period unconditional default rate is, the `invariant default profile' 
approach~\eqref{eq:defprofile.invariant} could be preferable for the forecast period 
despite its only moderate performance in our numerical examples.


\subsection{Case 2: No unconditional rating profile but unconditional PD given}
\label{se:case3}

In this case, we assume that a forecast unconditional PD $0 < p_1 < 1$ is given 
that is likely to differ from 
the estimation period unconditional PD. But the unconditional current rating profile
is assumed not to be known. This would typically be the case if a rating model was
newly developed and it was not possible to rate all the borrowers in the portfolio
in one big-bang effort. The new ratings would then only become available in the course of
the regular annual rating process. This is clearly suboptimal, in particular with a
view on the validation of the new rating model, but sometimes unavoidable due to limitation
of resources.

In this situation, proposition~\ref{pr:joint} suggests the assumption 
that both conditional rating profiles are invariant
as the only possibility to infer a full model of a borrower's beginning of the period
rating and end of the period state of solvency. \\
\textbf{Invariant conditional profiles:}
\begin{equation}\label{eq:condprofiles.invariant}
	\begin{split}
	\Pr_1[X=x\,|\,D] & \ = \ \Pr_0[X=x\,|\,D], \quad x=1, \ldots, k, \quad \text{and}\\
	\Pr_1[X=x\,|\,N] & \ = \ \Pr_0[X=x\,|\,N], \quad x=1, \ldots, k.
	\end{split}
\end{equation}
Note that \eqref{eq:condprofiles.invariant} is a stronger assumption than \eqref{eq:AR.invariant}
because \eqref{eq:AR.invariant} is implied by \eqref{eq:condprofiles.invariant}.

If, however, it is sufficient to obtain an estimate of the forecast period PD curve
then it is solely the estimation period likelihood ratio $x\mapsto \lambda_0(x)$ 
that one needs to know in addition
to the unconditional PD $p_1$. Formally, the assumption of an \textbf{invariant
likelihood ratio} is used here:
\begin{equation}\label{eq:likratio.invariant}
	\lambda_1(x)\ =\ \lambda_0(x), \quad x=1, \ldots, k.
\end{equation}
From
equation~\eqref{eq:pd.likratio} it follows that we can then calculate the PD curve as follows:
\begin{equation}\label{eq:condPD.1}
			\Pr_1[D\,|\,X = x] \ =\ \frac{p_1}{p_1 +
		(1-p_1)\,\lambda_0(x)}, \quad x = 1, \ldots, k.
\end{equation}
\citet[][theorem~2]{Elkan01} stated this observation
as `change in base rate' theorem. It is also often mentioned in the specific 
context of logistic regression \citep[see, for instance][section 6.2]{Cramer2003}.


\subsection{Case 3: Unconditional rating profile but no unconditional PD given}
\label{se:case2}

In this case, the unconditional rating profile $\Pr_1[X=x]$, $x=1, \ldots, k$ 
can directly be observed at the beginning of the forecast period.
Like in case~1, we consider several approaches to prediction in the forecast period that
may lead to proper models for the forecast period:
\begin{itemize}
	\item \textbf{Invariant PD curve.} Assume that the PD curve is invariant, i.e.\ the following equation holds:
	\begin{equation}\label{eq:PDcurve.invariant}	
	\Pr_1[D\,|\,X = x]  \ =\ \Pr_0[D\,|\,X = x], \quad x=1, \ldots, k. 
	\end{equation}	
	\item \textbf{Invariant conditional profiles.} Assume that both conditional rating profiles are invariant, i.e.\
	assumption~\eqref{eq:condprofiles.invariant}.
	\item \textbf{Invariant likelihood ratio.} Assume that the likelihood ratio is invariant, i.e.\ \eqref{eq:likratio.invariant} holds.
	Note that \eqref{eq:likratio.invariant} is implied by \eqref{eq:condprofiles.invariant} and, hence,
	is a weaker assumption.
\end{itemize} 

\subsubsection{On assumption \eqref{eq:PDcurve.invariant}} As mentioned in section~\ref{se:invariance},
it might not be the best idea to work under this assumption because there is a risk to `contaminate'
the forecast with the estimation period unconditional default rate $p_0$. However,
by proposition~\ref{pr:joint} the combination of unconditional rating profile with
any PD curve creates a unique proper model of a borrower's beginning of the period rating 
and end of the period state of solvency.
In particular, by \eqref{eq:uncondPD} this approach implies a forecast of the
unconditional default rate in the forecast period:
\begin{equation}
	p_1  \ = \ \sum_{x=1}^k \Pr_0[D\,|\,X = x]\,\Pr_1[X = x].
\end{equation}

\subsubsection{On assumption \eqref{eq:condprofiles.invariant}} Equation \eqref{eq:unconditional}
functions here as a constraint. The unknown unconditional PD $p_1$ and the two conditional profiles
therefore must satisfy
\begin{subequations}
\begin{equation}\label{eq:unconditional.constraint}
	\Pr_1[X = x] \ =\ p_1\,\Pr_0[X = x\,|\,D] + (1-p_1)\,\Pr_0[X = x\,|\,N], \quad x=1, \ldots, k.
\end{equation}
Hence, as all three profiles $\Pr_1[X = x]$, $\Pr_0[X = x\,|\,D]$, and $\Pr_0[X = x\,|\,N]$ are known,
we have $k$ equations for the one unknown $p_1$.
In general, it seems unlikely that all the $k$ equations can be simultaneously satisfied if only one 
variable can be freely chosen. However, we can try and compute a best fit by solving the following
least squares optimisation problem:
\begin{equation}\label{eq:leastsquares}
\begin{split}
		p_1^\ast & \ = \ \arg\min\limits_{p_1\in[0,1]} \sum_{x=1}^k \bigl(\Pr_1[X = x] - p_1\,\Pr_0[X = x\,|\,D] - 
			(1-p_1)\,\Pr_0[X = x\,|\,N]\bigr)^2 \\
	\Rightarrow\ p_1^\ast & \ = \ \sum_{x=1}^k \frac{\bigl(\Pr_1[X = x]-\Pr_0[X = x\,|\,N]\bigr)\,
	\bigl(\Pr_0[X = x\,|\,D]-\Pr_0[X = x\,|\,N]\bigr)}{\bigl(\Pr_0[X = x\,|\,D]-\Pr_0[X = x\,|\,N]\bigr)^2}.
\end{split}
\end{equation}
\end{subequations}
Observation \eqref{eq:leastsquares} is interesting because it indicates a technique to extract
a forecast of the unconditional PD from the unconditional rating profile at the beginning of the
forecast period that also avoids the contamination issue observed for the invariant PD curve assumption. 
It should be checked whether the forecast PD $p_1^\ast$ is indeed in line
with the profile $x\mapsto \Pr_1[X = x]$. This can readily be done because 
with $p_1^\ast$ from \eqref{eq:leastsquares} we obtain an implied unconditional rating profile
\begin{equation}
	\Pr_1^\ast[X = x] \ =\ p_1^\ast\,\Pr_0[X = x\,|\,D] + (1-p_1^\ast)\,\Pr_0[X = x\,|\,N], \quad x=1, \ldots, k.
\end{equation}
This can be $\chi^2$-tested against the grade-level frequencies of borrowers at the beginning
of the forecast period. If the hypothesis that $x \mapsto \Pr_1[X = x]$ is just a random
realisation of $x \mapsto \Pr_1^\ast[X = x]$ cannot be rejected we can proceed to predict the
PD curve on the basis of $x \mapsto \Pr_1^\ast[X = x]$ by using \eqref{eq:def.profile}:
\begin{equation}
	\Pr_1[D\,|\,X = x] \ =\ \frac{p_1^\ast\,\Pr_0[X = x\,|\,D]}{\Pr_1^\ast[X = x]}, \quad x=1, \ldots, k.
\end{equation}
The optimisation problem \eqref{eq:leastsquares} is convenient for deriving a forecast of $p_1$ from
the unconditional rating profile because it yields a closed-form solution. In principle, there is
no reason why the least squares should not be replaced with a -- say -- least absolute value 
optimisation. This would result in a slightly different forecast of $p_1$.  However, as we 
will check the appropriateness of the $p_1$ forecast by applying a $\chi^2$ test as
mentioned in section~\ref{se:observations}, it seems natural to also look at the variant of
\eqref{eq:leastsquares} where the $\chi^2$ statistic is directly minimised.  
It is easy to show that this minimisation problem
is well-posed and has a unique solution.

\subsubsection{On assumption \eqref{eq:likratio.invariant}} Like for assumption~\eqref{eq:condprofiles.invariant},
it is not a priori clear that a proper model of a borrower's rating profile and 
solvency state can be based on the unconditional profile $x\mapsto\Pr_1[X = x]$ and the 
likelihood ratio $\lambda_0(x)$. The necessary and sufficient condition for the likelihood
ratio to match the rating profile is provided in equation~\eqref{eq:condition} of theorem~\ref{th:unique},
with $\pi_x = \Pr_1[X = x]$ and $\lambda(x) = \lambda_0(x)$. 

If condition \eqref{eq:condition} is satisfied then proposition~\ref{pr:full} implies that
there is a unique model of a borrower's rating and solvency state
with characteristics $\Pr_1[X = x]$ and $\lambda_0(x)$. The unconditional PD in this
model is determined as the unique solution $p_1$ of equation~\eqref{eq:eqn}, and 
we can calculate the PD curve by \eqref{eq:condPD.1}.
\begin{table}[t!p]
\caption{Forecasts of 2010 and 2011 S\&P unconditional default rates
and p-values for goodness of fit tests of 2010 and 2011 unconditional rating profiles. 
The forecast approaches are described in section~\ref{se:case2}.}
\label{tab:p0.forecast}
\begin{center}
\begin{tabular}{|l||c|c||c|c|}
\hline
Forecast for & \multicolumn{2}{c||}{2010} & \multicolumn{2}{c|}{2011} \\ \hline
Observed default rate & \multicolumn{2}{c||}{1.14\%} & \multicolumn{2}{c|}{0.75\%} \\ \hline\hline
 & Forecast DR & p-value & Forecast DR & p-value \\ \hline \hline
Invariant PDs \eqref{eq:PDcurve.invariant} & 4.32\% & Exact & 3.75\% & Exact  \\ \hline
Least squares \eqref{eq:leastsquares} & 4.80\% & 0.0037 & 3.50\% & $< 10^{-10}$ \\ \hline
Least $\chi^2$ & 5.35\% & 0.0051 & 2.84\% & $< 10^{-10}$   \\ \hline
Invariant LR \eqref{eq:likratio.invariant} & 5.38\% & Exact & 2.79\% & Exact  \\ \hline
\end{tabular}
\end{center}
\end{table}

\subsubsection{Summary of section~\ref{se:case2}}	Table~\ref{tab:p0.forecast} on page~\pageref{tab:p0.forecast}
displays
some forecast results that were calculated with the approaches described
in this section. Forecast values for the 2010 and 2011 
S\&P unconditional default rates are presented together with
assessments of the goodness of fit of the actual 
unconditional rating profiles by the implied or assumed
unconditional rating profiles. It is immediately clear from table~\ref{tab:p0.forecast}
that the forecasts of the unconditional default rates are much too
high in all cases. That is why we did not bother to show
the grade-level forecast default rates or any other model characteristics as the
fit would have been equally poor.

We have argued above that assumption~\eqref{eq:PDcurve.invariant} is 
suboptimal for risking undesirable impact on the forecast of the estimation period
unconditional default rate. Assumption~\eqref{eq:likratio.invariant} 
is the most promising of the three assumptions we have explored because it guarantees
exact fit of the unconditional rating profile and avoids contamination of
the forecast. Assumption~\eqref{eq:condprofiles.invariant}
is stronger than \eqref{eq:likratio.invariant} because it implies \eqref{eq:likratio.invariant}.
 In principle, assumption~\eqref{eq:condprofiles.invariant}
will hardly ever provide a proper model because it is rather unlikely that the
overdetermined equation~\eqref{eq:unconditional.constraint} has an exact solution.
By \eqref{eq:leastsquares} or minimisation of the $\chi^2$ Pearson statistic, however, 
we could try and determine an approximate fit that could turn
out to be statistically indistinguishable from the rating profile at the beginning
of the forecast period -- which would make assumption~\eqref{eq:condprofiles.invariant} 
a viable approach, too. 

\begin{table}[t!p]
\caption{S\&P grade-level smoothed likelihood
ratios (defined by \eqref{eq:lik.ratio}) for
corporates in 2009, 2010 and 2011. Source:
Own calculations.}
\label{tab:lik.ratios}
\begin{center}
\begin{tabular}{|l||c|c|c|}
\hline
Rating grade & 2009 & 2010 & 2011  \\ \hline \hline
AAA & 1,501.55 & 53,720.74 & 20,221.19  \\ \hline
AA+ & 715.28 & 20,477.36 & 7,734.58  \\ \hline
AA & 353.69 & 8,583.13 & 3,410.45  \\ \hline
AA- & 167.12 & 3,099.52 & 1,365.00  \\ \hline
A+ & 88.18 & 1,161.49 & 547.00  \\ \hline
A & 45.53 & 436.37 & 227.20 \\ \hline
A- & 23.98 & 177.13 & 100.78  \\ \hline
BBB+ & 13.89 & 83.35 & 51.03 \\ \hline
BBB & 8.37 & 40.00 & 27.65 \\ \hline
BBB- & 5.17 & 19.85 & 15.00 \\ \hline
BB+ & 3.61 & 12.23 & 9.58 \\ \hline
BB & 2.70 & 8.28 & 6.74  \\ \hline
BB- & 1.78 & 4.93 & 4.24 \\ \hline
B+ & 1.01 & 2.46 & 2.19  \\ \hline
B & 0.48 & 0.96 & 0.79  \\ \hline
B- & 0.17 & 0.27 & 0.20  \\ \hline
CCC-C & 0.04 & 0.05 & 0.04  \\ \hline
\end{tabular}
\end{center}
\end{table}

As `contamination' by the 2009 unconditional default rate is prevented under 
assumptions~\eqref{eq:condprofiles.invariant} and \eqref{eq:likratio.invariant}, it is interesting
to speculate why the implied default rate forecasts are so poor nonetheless. 
The natural conclusion is that the assumptions are simply wrong for the S\&P data. 
Indeed, as table~\ref{tab:lik.ratios} on page~\pageref{tab:lik.ratios} demonstrates for the likelihood ratio, with 
hindsight it is clear that the
invariance assumptions made in this sections do not hold.
An alternative and complementary explanation could however be that the S\&P ratings made in 2009 and 2010
were over-pessimistic and for this reason generate too high default rate forecasts. This 
explanation is supported by the observation that in 2009 the downgrade-to-upgrade
ratio for the S\&P corporate ratings was 3.99 \citep[][table~6]{S&P2012} -- which
could presumably not even be compensated by the 2010 downgrade-to-upgrade
ratio of 0.74. Possibly, the truth is a mixture of these two explanations.


\subsection{Case 4: No unconditional rating profile and no unconditional PD given}
\label{se:case1}

From a risk management point of view it is undesirable to have
no current data at all. In a stable economic environment, this approach might
be justifiable nonetheless. One could assume
that the model from the estimation period works without any adaptations 
also for the forecast period. Of course, at the end of the forecast period,
we can then backtest the
default profile from the observation period against the grade-level default frequencies
observed. Formally, the assumption made in case~4
may be described by \eqref{eq:PDcurve.invariant} and
\begin{equation}
	\Pr_1[X=x]  \ = \ \Pr_0[X=x], \quad x=1, \ldots, k.
\end{equation}
Table \ref{tab:all}, combined with table~\ref{tab:pd.curves2009}, indicates that
it would not have been a good idea to try and predict the grade-level S\&P default rates
of 2010 and 2011 with the PD curve from 2009.
Alternatively, one might
try and come up with plausible assumptions on the forecast period unconditional rating profile or
unconditional PD -- which would bring us back into case 1, case 2, or case 3.


\begin{table}[t!p]
\caption{Moody's reported rating frequencies and one-year default rates for 1986.
Source: \citet[][Exhibit 41]{Moodys2013}. ``Def rate'' stands for ``default rate''.}
\label{tab:MoodysExample}
\begin{center}
\begin{tabular}{|l||c|c|c|c|c|}
\hline
Rating grade & Issuers & Def~rate & Implied defaults & Rounded defaults & `Rounded' def rate  \\ \hline \hline
Aaa & 108 & 0.00\% & 0 & 0 & 0.00\% \\ \hline
Aa  & 290 & 0.00\% & 0 & 0 & 0.00\% \\ \hline
A & 569 & 0.00\% & 0 & 0 & 0.00\% \\ \hline
Baa & 307 & 1.34\% & 4.11 & 4 & 1.30\% \\ \hline
Ba  & 357 & 2.87\% & 10.25 & 10 & 2.80\% \\ \hline
B & 187 & 11.57\% & 21.63 & 22 & 11.76\% \\ \hline
Caa-C & 10 & 22.22\% & 2.22 & 2 & 20.00\% \\ \hline
\end{tabular}
\end{center}
\end{table}

\section{Backtest}
\label{se:backtest}

In this section, we describe a backtest of the observations from section~\ref{se:case4} on a relatively long time series of
rating and default data for the years 1986 to 2012, as published\footnote{%
We discard the years 1970 to 1985 from the dataset because only from 1986 on there were at least 
10 issuers in each rating grade at the beginning of the year. Working with rating frequencies of less
than 10 would risk to make the results over-sensitive to random variation.}
 in \citet[][Exhibit 41]{Moodys2013}.
As mentioned in section~\ref{se:data}, this data is not optimal for the purpose of this paper. 
Table~\ref{tab:MoodysExample} illustrates the issue at the example of the data for the year 1986. 
The first three columns of table~\ref{tab:MoodysExample} have been extracted from \citet[][Exhibit 41]{Moodys2013}.
The fourth column `Implied defaults' has been determined by element-wise multiplication of the second and
third columns. The fact that the entries of the fourth column are not even approximately integers indicates that the default
rates from the third column were computed by a non-trivial method that involved information which is
not presented in \citet{Moodys2013}. However, in order to be able to apply the $\chi^2$ test 
for the goodness of fit of our estimates we need integer default numbers. In the following we 
adopt the obvious solution to this
problem by making use of rounded values as shown in the fifth column of table~\ref{tab:MoodysExample}.
A minor corruption of the data as shown in the sixth column
of the table is the price to pay for this solution. 

We repeat the calculations from section~\ref{se:case4} on the Moody's dataset but for lack of space do not present the detailed 
results. Table~\ref{tab:Result1986} illustrates the results of the calculations with the example of the grade-level 
default rates forecast for 1987 based on observations in 1986 (see table~\ref{tab:MoodysExample}). We use again the 
p-values of the $\chi^2$ tests to compare the goodness of fit achieved by the four different calibration methods. 
A method with a higher p-value is considered a better fit because the risk of a wrong decision by rejecting the resulting
calibration (i.e.~100\% minus p-value) is lower for such a method. 

\begin{table}[t!p]
\caption{1987 grade-level forecast default rates for
Moody's corporates ratings based on observations in 1986. P-values
are for the $\chi^2$-tests of the implied default profiles.
All values but the ranks are in \%.}
\label{tab:Result1986}
\begin{center}
\begin{tabular}{|l||c|c|c|c|c|c|}
\hline
 Rating  & Default & Invariant default & Invariant AR & Scaled PDs & Scaled likelihood \\ 
 grade &  rate & profile \eqref{eq:defprofile.invariant} & \eqref{eq:AR.invariant} & \eqref{eq:PDcurve.scaled} & 
ratio  \eqref{eq:likratio.scaled} \\ \hline\hline 
Aaa & 0 & 0.0099 & 0.0035 & 0.0039 & 0.0037 \\ \hline
Aa & 0 & 0.0446 & 0.0204 & 0.0222 & 0.0213 \\ \hline
A & 0 & 0.1732 & 0.0997 & 0.1183 & 0.1132 \\ \hline
Baa & 0 & 0.51 & 0.3519 & 0.4514 & 0.4323 \\ \hline
Ba & 2.8139 & 1.4717 & 1.2083 & 1.4611 & 1.4047 \\ \hline
B & 6.6667 & 6.7539 & 7.1427 & 7.2558 & 7.1365 \\ \hline
Caa-C & 20 & 51.6781 & 63.4415 & 44.2962 & 51.0725 \\ \hline\hline
P-value & Exact & 8.7716 & 7.5917 & 13.2315 & 10.5396 \\ \hline
Rank & & 2 & 1 & 4 & 3 \\ \hline
\end{tabular}
\end{center}
\end{table}
For a comparison of goodness of fit on one sample, inspecting the p-values is appropriate for ranking the different calibration
methods. This is, however, not the case if the comparison involves several samples as it does
in our backtesting exercise on the Moody's data. The issue with this is the sample-size dependence of the p-values.
On small samples, p-values are usually higher because the random variation is stronger. Hence, in years
with low default rates one would in general expect higher p-values than in years with high default rates.
As a consequence, it does not make sense to directly compare p-values that were calculated for different years.

What we can compare across several years, however, is the ranking of the different calibration methods. The last
row of table~\ref{tab:Result1986} show the ranks of the four calibration methods with regard to forecasting the
grade-level default rates for 1987. For that year, `scaled PDs' is best with the highest p-value and hence 
receives rank~4 while `invariant default profile' is worst and receives rank~1. We determined such rankings for
all the 26 years from 1987 to 2012 (each forecast was based on observations from the previous year) 
and then calculated the average rank for each of the four calibration methods.
Results are shown in table~\ref{tab:AverageRanks}.

\begin{table}[t!p]
\caption{Average ranks of calibration methods 
with respect to their $\chi^2$-test p-values for the years 1987 to 2012.}
\label{tab:AverageRanks}
\begin{center}
\begin{tabular}{|l||c|c|c|c|}
\hline
 Approach & Invariant default & Invariant AR & Scaled PDs & Scaled likelihood \\ 
 & profile \eqref{eq:defprofile.invariant} & \eqref{eq:AR.invariant} & \eqref{eq:PDcurve.scaled} & 
ratio  \eqref{eq:likratio.scaled} \\ \hline 
Average Rank & 2.58 & 2.05  & 2.61 &   2.76  \\ \hline
\end{tabular}
\end{center}
\end{table}

According to table~\ref{tab:AverageRanks} the `scaled likelihood ratio' approach performs best on average,
followed by the `scaled PDs' and `invariant default profile' approaches (with little difference), while 
the average performance of `invariant accuracy ratio' is worst. Hence, when compared with the observations from
section~\ref{se:case4}, the `scaled PDs' and `invariant accuracy ratio' approaches have swapped ranking positions.
The poor performance of `invariant accuracy ratio' in the backtest could be a sign that the underlying assumption 
of a constant accuracy ratio over time is simply not right\footnote{%
An anonymous referee pointed out that there is a strong negative correlation between unconditional
default rate and accuracy ratio over time. Indeed, for the data from \citet[][Exhibit 41]{Moodys2013} considered here
the rank correlation of default rate and accuracy ratio for the years 1986 to 2012 is -64.2\%.}.
The stronger than expected performance of the `scaled PDs' approach could be owed to its conceptual
similarity to the strong performing `scaled likelihood ratio'.


\section{Conclusions}
\label{se:conclusions}

Accurate (re-)calibration of a rating model requires careful
consideration of a number of questions that include, in particular,
the question of which model components can be assumed to be invariant
between the estimation period of the model and the forecast period.
Looking at PD curve calibration as a problem of forecasting rating-grade level
default rates, we have discussed a model framework that is suitable
for the description of a variety of different forecasting approaches.

We have then proceeded to present a number of PD curve calibration approaches
and explored the conditions under which the approaches are fit for purpose. 
We have tested the approaches introduced by applying them to publicly
available datasets of S\&P and Moody's rating and default statistics that can be
considered typical for the scope of application of the approaches.

One negative and one positive finding are the main results of our
considerations:
\begin{itemize}
	\item The popular `scaled PDs' approach for 
	(re-)calibrating a rating model to a different target unconditional PD
	is not likely to deliver the best calibration results because it implicitly mixes
	up the unconditional PD of the estimation period and the target PD.
	\item As shown by example, the `scaled likelihood ratio' approach
	to PD curve calibration
	avoids mixing up the
	unconditional PDs from the estimation and the forecast periods and,
	on average, performs better than `scaled PDs' and other approaches
	discussed in the paper. `Scaled likelihood ratio' is, therefore, 
	a promising alternative to `scaled PDs'. 
\end{itemize}


\appendix
\section{Appendix}
\label{se:app}

In this paper, we apply quasi moment matching (QMM) as suggested by \citet{Tasche2009a} for the smoothing of PD curves. 
QMM requires the numerical solution of a two-dimensional system of non-linear equations. The solution of such an equation system in general is much facilitated if a meaningful initial guess of the solution can be provided. 
The binormal model we discuss in the following subsection delivers such a guess. In addition, the binormal model provides the main motivation of the QMM technique. In subsection~\ref{se:QMM} we describe the QMM technique itself.

\subsection{The binormal model with equal variances}
\label{se:binormal}

Formally, the binormal model with equal variances is based on the following assumption.
\begin{assumption}\label{as:binormal}
$X$ denotes the continuous score of a borrower at the beginning
of the observation period.
\begin{itemize}
	\item The distribution of $X$ conditional on the event $D$ (the borrower defaults during the observation period) is normal with mean $\mu_D$ and variance $\sigma^2 > 0$.
	\item The distribution of $X$ conditional on the event $N$ (the borrower remains solvent during the whole observation period) is normal with mean $\mu_N > \mu_D$ and variance $\sigma^2 > 0$.
	\item $p \in (0,1)$ is the borrower's unconditional PD (i.e.\ the unconditional 
	probability that the borrower defaults during the observation period).
\end{itemize}
\end{assumption} 
Denote by $f_D$ and $f_N$ respectively the conditional densities of
the binormal score $X$. Hence by assumption~\ref{as:binormal} we have
\begin{equation}
	\begin{split}
	f_D(x) & = \frac{1}{\sqrt{2\pi}\,\sigma}\,\exp\bigl(-\frac{(x-\mu_D)^2}{2\,\sigma^2}\bigr),\\
	f_N(x) & = \frac{1}{\sqrt{2\pi}\,\sigma}\,\exp\bigl(-\frac{(x-\mu_N)^2}{2\,\sigma^2}\bigr).	
	\end{split}
\end{equation}
In the continuous case specified by assumption~\ref{as:binormal}, Bayes' formula implies a PD curve 
$x\mapsto \Pr[D\,|\,X=x]$ similar to the discrete formula \eqref{eq:pd_discrete}:
\begin{subequations}
\begin{align}
\Pr[D\,|\,X=x] & = 
		\frac{p\,f_D(x)}{p\,f_D(x) + (1-p)\,f_N(x)}	\label{eq:key}\\
		& = \frac{1}{1+\exp(\alpha+\beta\,x)},\label{eq:logit}\\
\alpha & = \frac{\mu_D^2-\mu_N^2}{2\,\sigma^2} + \log\left(\frac{1-p}{p}\right),\label{eq:alpha}\\
	\beta & = \frac{\mu_N-\mu_D}{\sigma^2}.\label{eq:beta}
\end{align}
\end{subequations}
Note that from \eqref{eq:logit} it follows that
\begin{equation}\label{eq:slope}
	\frac{d\,\Pr[D\,|\,X=x]}{d\,x} \ = \ - \beta\,\Pr[D\,|\,X=x]\,
	\bigl(1-\Pr[D\,|\,X=x]\bigr).
\end{equation}
Hence the absolute value of the slope of the PD curve \eqref{eq:logit} attains its maximum if and
only if $\Pr[D\,|\,X=x] = 1/2$ and then the maximum absolute 
slope is $\beta/4$.

Denote by $X_D$ and $X_N$ independent random variables with
$X_D \sim \mathcal{N}(\mu_D, \sigma)$ and $X_N \sim \mathcal{N}(\mu_N, \sigma)$.
Then, under assumption~\ref{as:binormal}, we also obtain a simple formula\footnote{%
$\Phi$ denotes the standard normal distribution function 
$\Phi(x) = \frac{1}{\sqrt{2\pi}} \int_{-\infty}^x e^{-1/2\,y^2}\,dy$.} for the discriminatory power
of the score $X$ if it is measured as \emph{accuracy ratio}
\citep[see, for instance,][section~3.1.1]{Tasche2009a}:
\begin{subequations}
\begin{align}
	\mathrm{AR} & = \Pr[X_D < X_N] - \Pr[X_D > X_N]\label{eq:generalAR}\\
	& = 2\,\Phi\left(\frac{\mu_N-\mu_D}{\sqrt{2}\,\sigma}\right)-1.\label{eq:AR.normal}
\end{align}
\end{subequations}
In addition, it is easy to show how the unconditional mean $\mu$ and variance $\tau^2$ of
the score $X$ can be described in terms of the means and variances of $X$ conditional on default and survival respectively:
\begin{subequations}
\begin{align}
\mu & = p\,\mu_D + (1-p)\,\mu_N, \label{eq:mu}\\
\tau^2 & = \sigma^2 + p\,(1-p)\,(\mu_D-\mu_N)^2.	\label{eq:tau}
\end{align}
\end{subequations}
A close inspection of equations \eqref{eq:AR.normal}, \eqref{eq:mu} and \eqref{eq:tau} shows that the conditional variance $\sigma^2$ and the conditional means $\mu_D$ and $\mu_N$ can be written as functions of
the unconditional mean, the unconditional variance and the accuracy ratio:
\begin{equation}\label{eq:FindCondPars}
\begin{split}
	c & = \sqrt{2}\,\Phi^{-1}\bigl(\tfrac{\mathrm{AR}+1}{2}\bigr), \\
	\sigma^2 & = \frac{\tau^2}{1+p\,(1-p)\,c^2}, \\
	\mu_N & = \mu + p\,\sigma\,c,\\
	\mu_D & = \mu - (1-p)\,\sigma\,c.
\end{split}
\end{equation}
From this, it follows by \eqref{eq:alpha} and \eqref{eq:beta} that also
the coefficients $\alpha$ and $\beta$ in \eqref{eq:logit} can be represented in terms of the unconditional mean $\mu$ of $X$, the 
unconditional variance $\tau^2$ of $X$, and the discriminatory power $\mathrm{AR}$ of $X$. In particular, we have the following representation of $\beta$ in terms of the accuracy ratio and the dispersion of the conditional score distributions:
\begin{equation}\label{eq:betaAR}
	\beta \ =\ \frac{\sqrt{2}\,\Phi^{-1}\bigl(\frac{\mathrm{AR}+1}2\bigr)}{\sigma}.
\end{equation}
These observations suggest the following three steps approach to identifying initial values for the QMM approach 
to PD curve smoothing:
\begin{enumerate}
	\item Calculate the mean $\mu$ and the standard deviation $\tau$ of the unconditional rating profile.
	\item Use $\mu$ and $\tau$ together with the unconditional PD $p$ and the accuracy ratio $\mathrm{AR}$ implied by the rating profile and the observed grade-level default rates to calculate the conditional 
 standard deviation $\sigma$ and the conditional means $\mu_D$ and $\mu_N$ according to \eqref{eq:FindCondPars}.
 \item Use equations \eqref{eq:alpha} and \eqref{eq:beta} to determine initial values for $\alpha$ and $\beta$.
\end{enumerate}
The initial values found by this approach will be the closer to the true values, the closer the conditional rating profiles are to normal distributions. 


\subsection{Quasi moment matching}
\label{se:QMM}

Equation \eqref{eq:key} shows that the unconditional PD has a direct primary
impact on the level of the PD curve. Equation \eqref{eq:betaAR} suggests that the AR
of a rating model has a similar impact on the maximum slope of the PD curve. The
two observations together suggest that in general a two-parameter PD curve can be
fitted to match given unconditional PD and AR.

It may be argued that for a suitably developed rating model based on carefully
selected risk factors, the associated PD curve must be monotonic for economic reasons.
Under the assumption that the PD curve is monotonic, \citet[][section~5.2]{Tasche2009a}
suggested the following robust version of the logistic curve \eqref{eq:logit} 
for fitting the PD curve:
\begin{equation}\label{eq:robustPD}
\begin{split}
	\Pr[D\,|\,X = x] & \ \approx \ \frac{1}
	{1 + \exp\bigl(\alpha + \beta\,\Phi^{-1}\bigl(F_N(x)\bigr)\bigr)}, \\
	F_N(x) & \ =\ \Pr[X \le x\,|\,N].
\end{split}
\end{equation}
This approach may be considered a variant of the ``sigmoid model'' suggested by
\citet{Platt99probabilisticoutputs}. The term $\Phi^{-1}\bigl(F_N(x)\bigr)$ in \eqref{eq:robustPD} transforms 
the in general non-normal distribution of the ratings conditional on survival into 
another distribution that is approximately normal even if the rating distribution 
is not continuous. However, in the discontinuous case $F_N(x) = 1$ may occur which
would entail $\Pr[D\,|\,X = x] = 0$. A suitable work-around to avoid this
is to replace the distribution function $F_N$ by the average $\widetilde{F}_N$ of $F_N$ and its
left-continuous version:
\begin{equation}\label{eq:rightleft}
	\widetilde{F}_N(x) \ = \ \frac{\Pr[X < x\,|\,N]+\Pr[X \le x\,|\,N]}2.
\end{equation}
Define, in addition to $F_N$, the distribution function $F_D$ of the rating variable $X$
conditional on default by
\begin{equation}
	F_D(x) \ =\ \Pr[X \le x\,|\,D].
\end{equation}
and denote by $X_D$ and $X_N$ independent random variables that are distributed 
according to $F_D$ and $F_N$ respectively.

For quantifying discriminatory power, we apply again the notion of \emph{accuracy ratio (AR)} as specified in \citet[][eq.~(3.28b)]{Tasche2009a}:
\begin{subequations}
\begin{equation}\label{eq:AR}
\begin{split}
	\mathrm{AR} &  = 2\,\Pr[X_D < X_N] + \Pr[X_D = X_N] - 1 \\
	& = \Pr[X_D < X_N] - \Pr[X_D > X_N]\\
	&= \int \Pr[X < x\,|\,D]\,d F_N(x) - \int \Pr[X < x\,|\,N]\,d F_D(x).
\end{split}
\end{equation}
See \citet[][section~2]{hand2001simple} for a discussion of why this definition of accuracy ratio (or
the related definition of the area under the ROC curve) is more expedient than the also common definition
in geometric terms. Definition \eqref{eq:AR} of AR takes an `ex post' perspective by assuming the 
obligors' states $D$ or $N$ at the end of the observation period are known and hence 
can be used for estimating the conditional (on default and survival respectively) 
rating distributions $F_D$  and $F_N$.

In the case where $X$ is realised as one of a finite number of rating grades $x=1, \ldots, k$,
the accuracy ratio can be calculated from the PD curve as follows:
\begin{equation}\label{eq:AR.discrete}
\begin{split}
	\mathrm{AR} & = \frac 1{p\,(1-p)} \Big(2 \sum_{x=1}^k \bigl(1-\Pr[D\,|\,X=x]\bigr)\,\Pr[X=x]\sum_{t=1}^{x-1} 
	\Pr[D\,|\,X=t]\,\Pr[X=t]\\
	&	\quad +
	 \sum_{x=1}^k \Pr[D\,|\,X=x]\,\bigl(1-\Pr[D\,|\,X=x]\bigr)\,\Pr[X=x]^2\Big) \ - 1,
\end{split}	 
\end{equation}
where $p$ stands for the unconditional PD as given by \eqref{eq:uncondPD}. 
\end{subequations}

Then \textbf{quasi moment matching}, for the purpose of this paper means the following procedure:
\begin{enumerate}
	\item Fix target values 
$p_{\mathrm{target}}$ and $\mathrm{AR}_{\mathrm{target}}$ for the unconditional portfolio PD and the 
accuracy ratio of the rating model.
\item Substitute $p_{\mathrm{target}}$ and $AR_{\mathrm{target}}$ for the left-hand sides of
equations \eqref{eq:uncondPD} and \eqref{eq:AR.discrete} respectively.
\item Represent $\Pr[D\,|\,X = x]$ in \eqref{eq:uncondPD} and \eqref{eq:AR.discrete}
by the robust logistic curve \eqref{eq:robustPD} (with $F_N$ replaced by $\widetilde{F}_N$).
Determine $\widetilde{F}_N$ by means of \eqref{eq:ndef.profile}
from the empirical unconditional rating profile, the grade-level default rates and
the unconditional default rate.
\item Choose initial values for the parameters $\alpha$ and $\beta$ according to 
\eqref{eq:alpha}, \eqref{eq:beta} and \eqref{eq:FindCondPars}, with
$\mu = \mathrm{E}\bigl[\Phi^{-1}\bigl(\widetilde{F}_N(X)\bigr)\bigr]$ and
$\tau^2 = \mathrm{var}\bigl[\Phi^{-1}\bigl(\widetilde{F}_N(X)\bigr)\bigr]$.
\item Solve numerically the equation system for $\alpha$ and $\beta$.
\end{enumerate}


\end{document}